\documentclass[twocolumn]{aastex63}

\expandafter\def\csname editcolor2\endcsname{blue}  
\expandafter\def\csname editcolor3\endcsname{orange} 
\expandafter\def\csname editcolor4\endcsname{violet}

\usepackage{soul}
\def\jing#1{{\colorcount=1\color{\csname editcolor\the\colorcount\endcsname}
{\ifturnoffone\else (Jing: #1)\ \fi}}}

\def\eric#1{{\colorcount=2\color{\csname editcolor\the\colorcount\endcsname}
{\ifturnofftwo\else\bf (Eric: #1)\ \fi}}}

\def\law#1{{\colorcount=3\color{\csname editcolor\the\colorcount\endcsname}
{\ifturnofftwo\else\bf (Lawrence: #1)\ \fi}}}

\def\chris#1{{\colorcount=3\color{\csname editcolor\the\colorcount\endcsname}
{\ifturnofftwo\else\bf (Chris A: #1)\ \fi}}}

\def\note#1{{\colorcount=4\color{\csname editcolor\the\colorcount\endcsname}
{\ifturnofftwo\else\bf (#1)\ \fi}}}

\usepackage{float}
\usepackage{bookmark} 
\bookmarksetup{numbered, open,}
\usepackage{enumitem}
\setlist[enumerate]{itemsep=0mm}
\usepackage{hyperref}
\usepackage{subfigure}
\usepackage{multirow}
\usepackage{xspace}
\usepackage{natbib}
\usepackage{xcolor}  
\usepackage{lineno} 
\usepackage{CJK} 

\usepackage{soul}
\usepackage{amsfonts}

\graphicspath{{./}{figures/}}

\newcommand{\declinerate}{$\Delta m_{15}(B)$}
\newcommand{\um}{$\mu$m}
\newcommand{\GPRthreshold}{0.2}

\received{November 11, 202}
\revised{XXX}
\accepted{XXX}
\submitjournal{ApJ}
\shorttitle{NIR spectral templates}
\shortauthors{Lu et al.}

\begin{document}

\title{Carnegie Supernova Project-II: Near-infrared spectral diversity and template of Type Ia Supernovae}

\correspondingauthor{Jing Lu}
\email{lujingeve158@gmail.com}

\author[0000-0002-3900-1452]{Jing Lu \begin{CJK*}{UTF8}{gbsn}(陆晶)\end{CJK*}}
\affil{Department of Physics, Florida State University, 77 Chieftan Way, Tallahassee, FL 32306, USA}

\author[0000-0003-1039-2928]{Eric Y. Hsiao \begin{CJK*}{UTF8}{bsmi}(蕭亦麒)\end{CJK*}}
\affil{Department of Physics, Florida State University, 77 Chieftan Way, Tallahassee, FL 32306, USA}

\author[0000-0003-2734-0796]{Mark M. Phillips}
\affil{Carnegie Observatories, Las Campanas Observatory, Colina El Pino, Casilla 601, Chile}

\author[0000-0003-4625-6629]{Christopher R. Burns}
\affil{The Observatories of the Carnegie Institution for Science, 813 Santa Barbara St., Pasadena, CA 91101, USA}

\author[0000-0002-5221-7557]{Chris Ashall}
\affil{Department of Physics, Virginia Tech, Blacksburg, VA 24061, USA}

\author[0000-0003-2535-3091]{Nidia Morrell}
\affil{Carnegie Observatories, Las Campanas Observatory, Colina El Pino, Casilla 601, Chile}

\author[0000-0002-3468-8558]{Lawrence Ng}
\affil{Department of Physics, Florida State University, 77 Chieftan Way, Tallahassee, FL 32306, USA}

\author[0000-0001-8367-7591]{Sahana Kumar}
\affil{Department of Physics, Florida State University, 77 Chieftan Way, Tallahassee, FL 32306, USA}

\author[0000-0002-9301-5302]{Melissa Shahbandeh}
\affil{Department of Physics and Astronomy, Johns Hopkins University, Baltimore, MD 21218, USA}
\affil{Space Telescope Science Institute, 3700 San Martin Drive, Baltimore, MD 21218, USA}

\author[0000-0002-4338-6586]{Peter Hoeflich}
\affil{Department of Physics, Florida State University, 77 Chieftan Way, Tallahassee, FL 32306, USA}

\author[0000-0001-5393-1608]{E. Baron}
\affiliation{Homer L. Dodge Department of Physics and Astronomy, University of Oklahoma, 440 W. Brooks, Rm 100, Norman, OK USA}
\affiliation{Hamburger Sternwarte, Gojenbergsweg 112, 21029 Hamburg, Germany}

\author[0000-0002-9413-4186]{Syed Uddin}
\affil{George P. and Cynthia Woods Mitchell Institute for Fundamental Physics and Astronomy, Department of Physics and Astronomy, Texas A\&M University, College Station, TX, 77843, USA}

\author[0000-0002-5571-1833]{Maximilian D. Stritzinger}
\affil{Department of Physics and Astronomy, Aarhus University, Ny Munkegade, DK-8000 Aarhus C, Denmark}

\author[0000-0002-8102-181X]{Nicholas B. Suntzeff}
\affil{George P. and Cynthia Woods Mitchell Institute for Fundamental Physics and Astronomy, Department of Physics and Astronomy, Texas A\&M University, College Station, TX, 77843, USA}

\author[0000-0003-0424-8719]{Charles Baltay}
\affil{Physics Department, Yale University, 217 Prospect Street, New Haven, CT 06511, USA}

\author[0000-0002-2806-5821]{Scott Davis}
\affil{Department of Physics, Florida State University, 77 Chieftan Way, Tallahassee, FL 32306, USA}

\author[0000-0002-0805-1908]{Tiara R. Diamond}
\affil{Private astronomer, tiaradiamond@gmail.com}

\author[0000-0001-5247-1486]{Gaston Folatelli}
\affil{Facultad de Ciencias Astron\'{o}micas y Geof\'{i}sicas, Universidad Nacional de La Plata, Instituto de Astrof\'{i}sica de La Plata (IALP), CONICET, Paseo del Bosque S/N, B1900FWA La Plata, Argentina}
\affil{Kavli Institute for the Physics and Mathematics of the Universe (WPI), The University of Tokyo, 5-1-5 Kashiwanoha, Kashiwa, Chiba 277-8583, Japan}

\author[0000-0003-3459-2270]{Francisco F\"{o}rster}
\affil{Millennium Institute of Astrophysics, Casilla 36-D,7591245, Santiago, Chile}
\affil{Departamento de Astronom\'{i}a, Universidad de Chile, Casilla 36-D, Santiago, Chile}

\author[0000-0002-2592-9612]{Jonathan Gagn\'{e}}
\affil{Plan\'etarium Rio Tinto Alcan, Espace pour la Vie, 4801 av. Pierre-de Coubertin, Montr\'eal, Qu\'ebec, Canada}
\affil{Carnegie Observatories, Las Campanas Observatory, Colina El Pino, Casilla 601, Chile}

\author[0000-0002-1296-6887]{Llu\'{i}s Galbany}
\affil{Institute of Space Sciences (ICE, CSIC), Campus UAB, Carrer de Can Magrans, s/n, E-08193 Barcelona, Spain}
\affil{Institut d'Estudis Espacials de Catalunya (IEEC), E-08034 Barcelona, Spain}

\author[0000-0002-8526-3963]{Christa Gall}
\affil{DARK, Niels Bohr Institute, University of Copenhagen, Jagtvej 128, 2200 Copenhagen, Denmark}

\author[0000-0001-9541-0317]{Santiago Gonz\'alez-Gait\'an}
\affiliation{CENTRA - Centro de Astrof\'isica e Gravita\c{c}\~{a}o, Instituto Superior T\'ecnico, Av. Rovisco Pais 1, 1049-001, Lisbon, Portugal}

\author{Simon Holmbo}
\affil{Department of Physics and Astronomy, Aarhus University, Ny Munkegade, DK-8000 Aarhus C, Denmark}

\author[0000-0002-1966-3942]{Robert P. Kirshner}
\affil{TMT International Observatory, 100 West Walnut Street, Pasadena, CA 91124}
\affil{California Institute of Technology, 1216 E. California Boulevard, Pasadena, CA 91125}

\author[0000-0002-6650-694X]{Kevin Krisciunas}
\affil{George P. and Cynthia Woods Mitchell Institute for Fundamental Physics and Astronomy, Department of Physics and Astronomy, Texas A\&M University, College Station, TX, 77843, USA}

\author{G. H. Marion}
\affil{Department of Astronomy, University of Texas, 1 University Station C1400, Austin, TX 78712, USA}

\author[0000-0002-4436-4661]{Saul Perlmutter}
\affil{Lawrence Berkeley National Laboratory, Department of Physics, 1 Cyclotron Road, Berkeley, CA 94720, USA}
\affil{Physics Department, University of California at Berkeley, Berkeley, CA 94720, USA}

\author[0000-0002-8041-8559]{Priscila J. Pessi}
\affil{Facultad de Ciencias Astron\'omicas y Geof\'isicas, Universidad Nacional de La Plata, Paseo del Bosque S/N, B1900FWA, La Plata, Argentina}

\author[0000-0001-6806-0673]{Anthony L. Piro}
\affil{The Observatories of the Carnegie Institution for Science, 813 Santa Barbara St., Pasadena, CA 91101, USA}

\author{David Rabinowitz}
\affil{Physics Department, Yale University, 217 Prospect Street, New Haven, CT 06511, USA}

\author[0000-0003-4501-8100]{Stuart D. Ryder}
\affil{School of Mathematical and Physical Sciences, Macquarie University, NSW 2109, Australia}
\affil{Astronomy, Astrophysics and Astrophotonics Research Centre, Macquarie University, Sydney, NSW 2109, Australia}

\author[0000-0003-4102-380X]{David J. Sand}
\affiliation{Department of Astronomy and Steward Observatory, University of Arizona, 933 N Cherry Avenue, Tucson, AZ 85719, USA}

\begin{abstract}
We present the largest and most homogeneous collection of near-infrared (NIR) spectra of Type Ia supernovae (SNe~Ia): 339 spectra of 98 individual SNe obtained as part of the Carnegie Supernova Project-II.
These spectra, obtained with the FIRE spectrograph on the 6.5~m Magellan Baade telescope, have a spectral range of 0.8--2.5~\um.
Using this sample, we explore the NIR spectral diversity of SNe~Ia and construct a template of spectral time series as a function of the light-curve-shape parameter, color stretch $s_{BV}$.
Principal component analysis is applied to characterize the diversity of the spectral features and reduce data dimensionality to a smaller subspace.
Gaussian process regression is then used to model the subspace dependence on phase and light-curve shape and the associated uncertainty. 
Our template is able to predict spectral variations that are correlated with $s_{BV}$, such as the hallmark NIR features: \ion{Mg}{2} at early times and the $H$-band break after peak. 
Using this template reduces the systematic uncertainties in K-corrections by $\sim$90\% compared to those from the Hsiao template.
These uncertainties, defined as the mean K-correction differences computed with the color-matched template and observed spectra, are on the level of $4\times10^{-4}$~mag on average.
This template can serve as the baseline spectral energy distribution for light-curve fitters and can identify peculiar spectral features that might point to compelling physics.
The results presented here will substantially improve future SN~Ia cosmological experiments, for both nearby and distant samples.
\end{abstract}



\section{Introduction} \label{sec:intro}

Type Ia supernovae (SNe~Ia) provide excellent luminosity distances thanks to their empirically standardizable properties, such as the correlations between the intrinsic peak luminosity and the light-curve shape \citep{Pskovskii1977,Phillips1993}, as well as the color \citep{Tripp1998}.
The comparison of the high-redshift SNe~Ia with the low-redshift sample led to the first evidence for cosmic acceleration \citep{Hamuy1996a, Riess1998,Perlmutter1999}.
By mapping out the expansion history of the Universe, SNe~Ia have become an essential tool for understanding the nature of dark energy \citep[e.g.][]{Riess2007,Suzuki2012,Scolnic2018,Brout2022}.

The precision of SN~Ia cosmology is now limited by systematic errors that cannot be reduced by simply observing more SNe.
These include uncertainties introduced by photometric calibration, evolutionary effects, host-galaxy environment, and uncertain extinction laws.
Observations in the near-infrared (NIR) offer a promising direction to achieve more accurate results for cosmology \citep[e.g.][]{BaroneNugent2012,Stanishev2018,Burns2018,Avelino2019,Jones2022} with several advantages compared to the optical. 

The NIR region is less affected by dust extinction, minimizing the reliance on uncertain dust laws \citep[e.g.][]{Cardelli1989,Krisciunas2000,Johansson2021}.
The bias toward low-reddening objects in high-redshift observations is also minimized in the NIR, reducing the systematic differences between the colors of low- and high-redshift SN~Ia samples.
Furthermore, both theory \citep{Kasen2006} and observations \citep{Krisciunas2004,WoodVasey2008,Mandel2009,Folatelli2010,Kattner2012} have shown that SNe~Ia have more uniform peak luminosities in the NIR.
Therefore, smaller light-curve-shape correction is needed to standardize normal SNe~Ia in the NIR  \citep[e.g.][]{Krisciunas2004,Dhawan2018a,Galbany2022}, which could reduce the systematic errors caused by potential evolutionary effects.

Due to the lack of sufficient NIR data, the majority of the empirical light-curve fitters currently available for SN cosmology rely on optical data only and have not yet included NIR coverage.
The only three light-curve models that cover NIR so far are \textit{SuperNovae in object-oriented Python} \citep[\texttt{SNooPy};][]{Burns2011,Burns2014}, the hierarchical Bayesian SED Model for SNe~Ia \citep[\texttt{BayeSN};][]{Mandel2009,Mandel2011,Mandel2022}, and \texttt{SALT3-NIR} \citep{Pierel2022_SALT3NIR}.

K-corrections are necessary to allow for the comparison of rest-frame magnitudes of objects at various redshifts \citep{Oke1968}.
Several efforts have been made to improve the accuracy of the K-corrections for SNe~Ia, such as the utilization of the cross-filter K-corrections for high-redshift SNe~Ia \citep{Kim1996} and the construction of time-series spectral templates \citep{Nugent2002,Hsiao2007} of the spectral energy distribution (SED). 
Uncertainties in the K-corrections, caused by intrinsic variations in the broadband colors and spectral features, directly affect the error budget of the distance measurements \citep[e.g.][]{Hsiao2007,Boldt2014}.

As the next generation of space telescopes such as the James Webb Space Telescope and the Nancy Grace Roman Space Telescope (RST) look further to the red, an accurate description of how the SN~Ia SED varies in the NIR is important for a wide range of applications.
It is particularly crucial for any NIR SN~Ia dark energy experiments, whether explicit K-correction calculations are needed or not.
For example, in the case of SALT3 \citep{Kenworthy2021}, an accurate underlying SED model is required to evade K-correction calculation.
However, current NIR spectral templates have room for improvements.
For example, the NIR region of the spectral template of \citet{Hsiao2007} and \citet{Hsiao2009} is only based on 52 NIR spectra of 30 SNe~Ia compiled by \citet{Marion2009}, a sample too small to adequately capture intrinsic variations.
Further complicating the matter, the features in the regions of strong telluric absorptions are completely obscured due to its narrow range of redshifts in this sample.

As in the optical, the NIR spectral features of SNe Ia are highly correlated with the light-curve shape, the primary parameter \citep{Hsiao2009}.
The correlation is especially evident in the $H$-band break \citep[e.g.][]{Hsiao2013,Ashall2019a,Ashall2019b}, the most prominent SN~Ia spectral feature in the NIR \citep{Kirshner73,Wheeler1998}.
This suggests that the SEDs can be accurately predicted given only the light-curve shape  \citep{Hsiao2007,Hsiao2009}.
Characterizing the NIR spectroscopic diversity of SNe~Ia became one of the main motivations for the second stage of the Carnegie Supernova Project \citep[CSP-II;][]{Phillips2019,Hsiao2019}.

CSP recognizes the potential of NIR observations and contributed high-precision observations of hundreds of SNe~Ia over the past roughly two decades. 
The first phase of CSP \citep[CSP-I; 2004-2009;][]{Hamuy2006} aimed to construct a data set of both optical and NIR light curves in well-understood and calibrated photometric systems \citep{Contreras2010,Stritzinger2011,Krisciunas2017} with telescopes located at Las Campanas Observatory.
A separate high-redshift ($0.1 < z < 0.7$) component of the project produced a NIR Hubble diagram in the rest-frame $i$ band and examined the potential of shifting SN~Ia cosmology toward the red \citep{Freedman2009}.

CSP-II (2011-2015) focused on building the low-redshift optical and NIR anchor with an unbiased sample of SNe~Ia in the smooth Hubble flow \citep{Phillips2019}.
A second component of the project aimed to obtain a statistically significant sample of NIR spectra \citep{Hsiao2019} for SNe of all types \citep[e.g.,][]{Davis2019,Shahbandeh2022}.
In collaboration with the Harvard-Smithsonian Center for Astrophysics (CfA) Supernova Group, a large sample was mainly acquired using the then newly commissioned Folded-port InfraRed Echellette \citep[FIRE;][]{Simcoe2013} mounted on the Magellan Baade telescope.
Over the course of the CSP-II campaign, 620 NIR spectra of 149 SNe~Ia were obtained.

The CSP-II sample is much larger than those in previous works (roughly an order of magnitude increase), and most observations are accompanied by accurate photometry.
The vast majority of the SNe~Ia have time-series NIR spectroscopic observations and complementary photometry, such that the phases of the spectra and the light-curve-shape parameters of the SNe~Ia can be accurately determined.
Another key improvement is in the regions of strong telluric absorption.
The high throughput of FIRE, in combination with the large aperture of Baade, provided an adequate signal to recover the SN flux in the telluric regions for most of our spectra.
Using this data set, we construct NIR time-series spectral templates as a function of phase and light-curve-shape parameter.

In this paper, we present the NIR spectra obtained by the CSP-II and a new set of NIR spectral templates constructed using this data set. 
Section~\ref{sec:sample} describes the selection criteria and  characteristics of the sample used.
In Section~\ref{sec:methods}, the methodology of the template construction is described in detail.
The properties of the template and its impact on cosmology are discussed in Section~\ref{sec:results}.
Finally, a summary is presented in Section~\ref{sec:conclusion}.
In this work, unless noted otherwise, the phase is relative to the time of the rest-frame $B$-band light-curve maximum.


\section{Sample}\label{sec:sample}

In this section, we describe the data set used in this work. 
The sample of NIR spectra comes entirely from the CSP-II \citep{Phillips2019,Hsiao2019}.
Approximately three-quarters of SN~Ia spectra were obtained using FIRE in the high-throughput prism mode.
In contrast with the majority of other NIR spectrographs, this configuration of FIRE consistently provides an adequate signal in the regions of heavy telluric absorption to enable telluric corrections.
Using the P$\alpha$ P-Cygni feature in our sample of type II SN (SN~II) spectra, the telluric corrections were shown to recover the feature to 10\% precision or better in 70\% of our sample \citep{Davis2019}.
The conventional notion that features in the telluric regions can only be obtained by space-based or airborne observatories is no longer correct.

The features in the telluric regions are crucial, as they will inevitably be shifted into observed filters when we study SNe~Ia at a range of redshifts.
This was the main difficulty in obtaining accurate NIR K-correction uncertainties in past studies \citep[e.g.,][]{Boldt2014}.
For this reason, as well as to ensure the homogeneity of the data set, we opted to only include FIRE spectra taken in the prism mode\footnote{Note that a few FIRE spectra obtained by the CSP-II were taken in the echelle mode.} for this work.
This amounts to an initial sample of 461 NIR spectra of 142 SNe~Ia which we narrowed through the selection criteria described below.
The spectra observed in FIRE prism mode have an approximate wavelength coverage from 0.8 to 2.5~$\mu$m, and resolutions of R $\sim$ 500, 450, and 300 in the $J$, $H$, and $K$ bands, respectively.
Details of the observations and data reduction can be found in \citet{Hsiao2019}.

\begin{figure}[tb!]
\centering
\includegraphics[width=0.98\columnwidth]{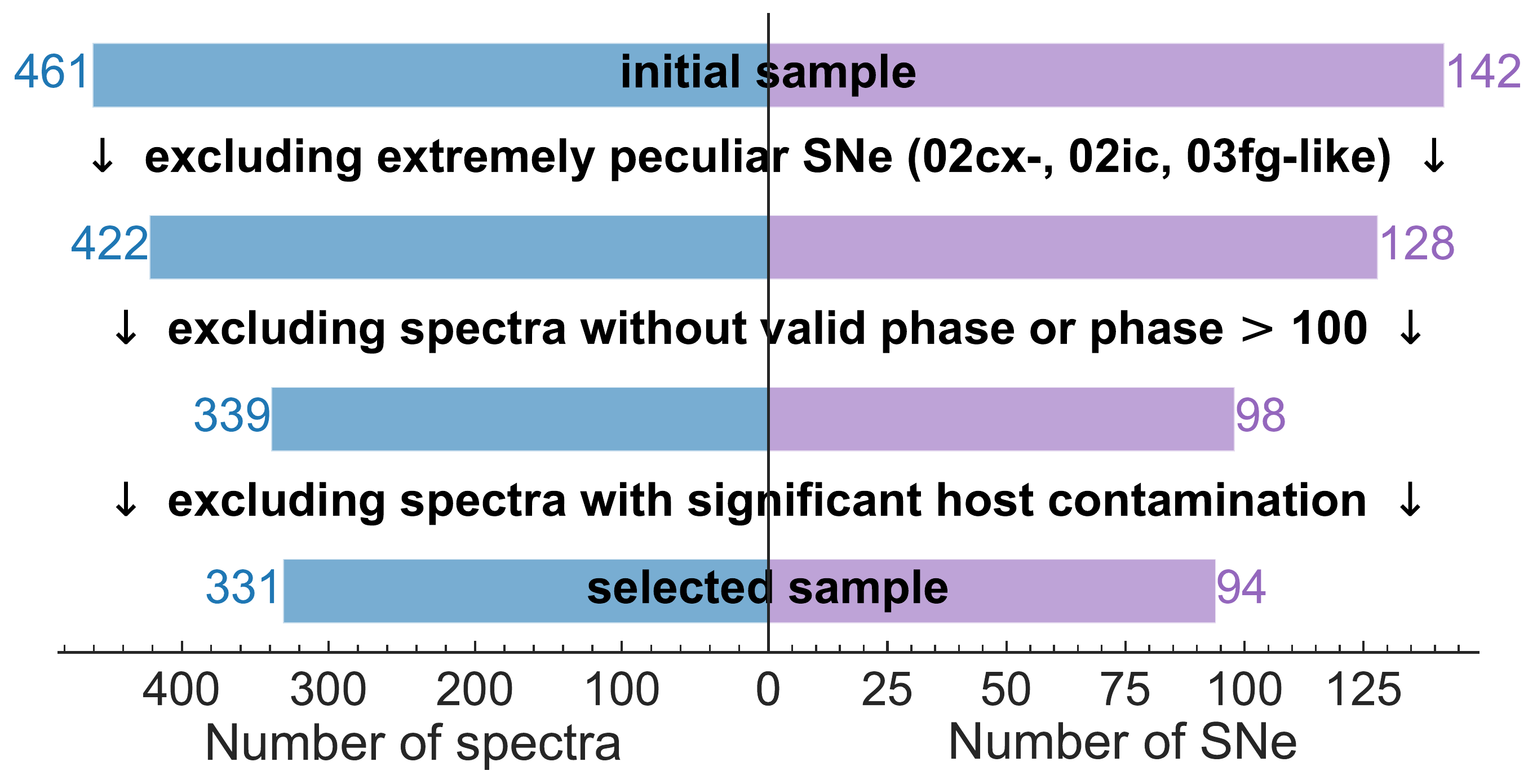} 
\caption{The number of spectra and SNe~Ia as each selection criterion is applied.}
\label{fig:sample_selection}
\end{figure}

\subsection{Selection Criteria} \label{subsec: sample selection}

With the goal of constructing spectral templates as a function of the color-stretch parameter \citep[$s_{BV}$;][]{Burns2014} and phase, the following selection criteria were applied in this order:

\begin{itemize}
\item Spectra of SNe~Ia that belong to one of the following peculiar subgroups: 02cx-like \citep{Li03}, 02ic-like \citep{Hamuy2003}, 03fg-like \citep{Howell2006}, and Ca-strong \citep{Galbany2019} were removed due to their peculiar spectral features.
This step excludes 39 spectra of 14 SNe.

\item Each SN must have a reliable time of the $B$-band maximum ($T_{\text{max}}^{B}$) and each spectrum must have a phase that is less than or equal to 100~days. 
This step further excludes 83 spectra of 30 SNe~Ia.

\item Each spectrum was visually inspected for significant host-galaxy contamination by comparing with spectra in a similar parameter space. 
This final step further excludes 8 spectra of 4 SNe.
\end{itemize}

Out of the initial 461 spectra of 142 SNe, 331 spectra of 94 SNe~Ia remain after the three-step sample cut.
Figure~\ref{fig:sample_selection} illustrates the change of sample size after applying each selection criterion.
All spectra that passed through the first two steps, 339 spectra of 98 SNe, are available in the CSP website\footnote{\url{https://csp.obs.carnegiescience.edu/data}}. 
A log of these spectra is listed in Appendix~\ref{appendix:NIR spec table}.

$T_{\text{max}}^{B}$ and the $s_{BV}$ of the sample SNe were obtained by fitting multiband CSP light curves using \texttt{SNooPy} with \texttt{max\_model}.
An arbitrary cutoff phase was set at 100~days relative to $B$-band maximum, well past the phase range relevant for light-curve fitting.

The peculiar SNe~Ia were identified based on their light curves and optical spectra.
Note that spectra of the slow-declining 91T-like \citep{Filippenko1992a,Phillips1992} and fast-declining 91bg-like \citep{Filippenko1992b,Leibundgut1993} SNe~Ia were kept in the sample intentionally for the following reasons:
1) Their NIR spectral features, while extreme, were shown to extend the properties of normal SNe~Ia \citep[e.g.,][]{Hoeflich2002, Hsiao2015,Phillips2022}.
2) Studies have shown that 91bg-like and fast-declining SNe~Ia in general are standardizable and do not impact Hubble constant measurements significantly \citep[e.g.,][]{Burns2014, Hoogendam2022,Phillips2022,Yang2022}.
3) The inclusion of these SNe will help broaden the $s_{BV}$ baseline since they are on the edges of the distribution.

Only 8 spectra show significant host contamination.
These are usually SNe~Ia situated close to the host galaxy core and taken at late phases when the background flux is comparable to or higher than the SN itself.
The effects are shallow spectral slope and more diluted spectral features especially toward the red end.
These were easy to identify through visual inspection and comparison to other spectra at similar phases.
They also showed up as outliers during the dimensionality reduction step (Section~\ref{sec:PCA}) if there were included.

\begin{figure*}[htb!]
\centering
\includegraphics[width=0.998\textwidth]{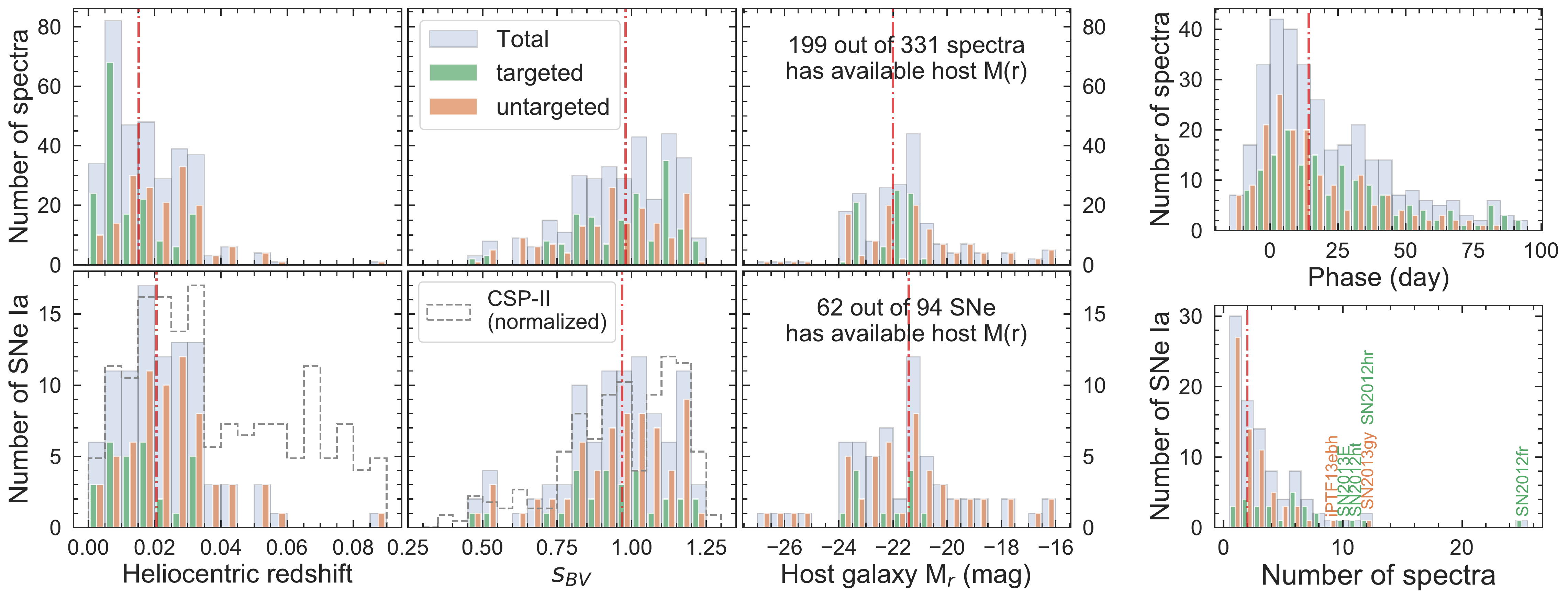} 
\caption{
Characteristics of the selected sample of NIR spectra. 
The top panels present the distributions of spectra in redshift, $s_{BV}$, host galaxy $r$-band absolute magnitude and phase; the bottom panels show the distributions of SNe in redshift, $s_{BV}$, host galaxy $r$-band absolute magnitude, as well as the number of spectra per SN.
The green and orange distributions represent SNe discovered by targeted and untargeted surveys, respectively.
The median value of each total distribution is indicated with a vertical dash-dotted line.
For comparison, the distributions (normalized to peak) of the full CSP-II SNe~Ia sample from \citet{Phillips2019} are plotted with dashed lines in the lower panels.
}
\label{fig:hists}
\end{figure*}

\subsection{Sample Characteristics} \label{subsec:sample_distribution} 

The final sample of spectra that passed the selection criteria (331 spectra of 94 SNe) cover a wide range of heliocentric redshift, $s_{BV}$, and phase (see Figure~\ref{fig:hists}).
A full range in $s_{BV}$ and phase is desirable as the goal here is to build a set of spectral templates as a function of $s_{BV}$ and phase.
A wide range in redshift is also useful as spectral features are shifted out of regions that are heavily affected by telluric absorption.
The sample SNe cover a heliocentric redshift range of $0.003 \le z \le 0.087$ with a median of 0.021 and a color stretch range of $0.48 \le s_{BV} \le 1.30$ with a median of 0.97.
The $r$-band absolute magnitudes of the host galaxies are obtained using the Z-PEG \citep{LeBorgne2002} SED software following the procedure described in \citet{Uddin2017}.

The majority of the SNe~Ia (70\%) in this sample were discovered by untargeted searches.
This is the result of the survey strategy of CSP-II, designed to avoid bias towards bright and large host galaxies \citep{Phillips2019}.  
As shown in the bottom panels of Figure~\ref{fig:hists}, SNe from untargeted searches extend out to higher redshifts and cover a wider range of the host galaxy absolute magnitude compared to those from targeted searches.
They also have a similar distribution in $s_{BV}$ except for a possible excess at the bright and slowly-declining end.
The normalized distributions of the full CSP-II SNe~Ia sample \citep[combined ``Physics'' and ``Cosmology'' samples from][]{Phillips2019} are also shown in Figure~\ref{fig:hists}.
The selected sample for the template is representative of the full CSP-II sample in terms of $s_{BV}$. 

\begin{figure*}[bt!]
\centering
\includegraphics[width=0.8\textwidth]{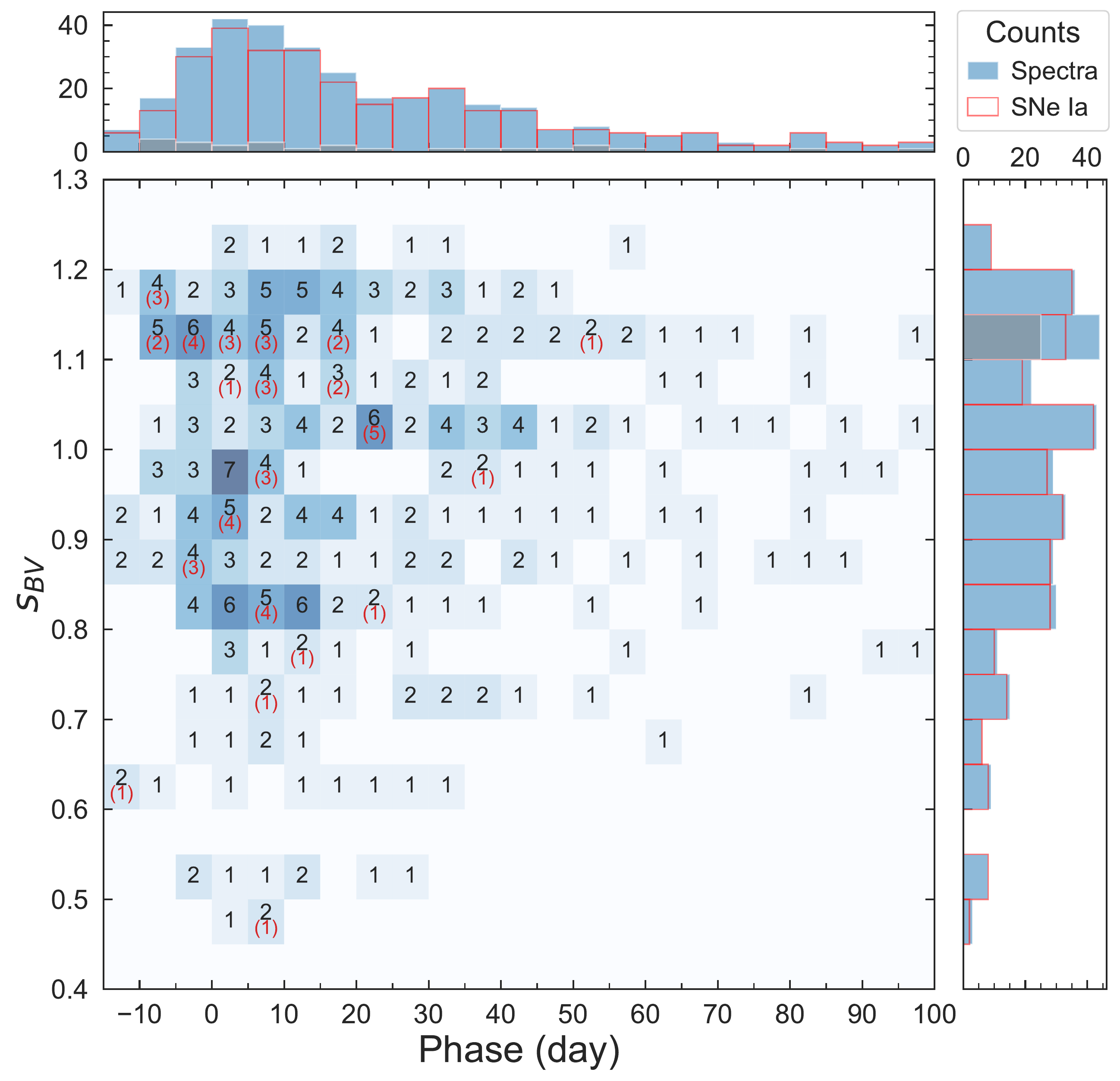} 
\caption{Two-dimensional histogram of the spectra/SNe in (phase, $s_{BV}$) with marginal distributions.
The bin size is 5~days for phase and 0.05 for $s_{BV}$.
Each non-empty bin has a black label to show the number of spectra, with an additional red label beneath indicating the number of SNe if there are more than 1 spectrum that belongs to the same SN.
The gray filled distribution represents the most well-sampled SN~2012fr in the selected data set.
}
\label{fig:counts_phase_sBV_grid}
\end{figure*}

Next, we examine the distribution of the sample spectra and SNe in phase$-s_{BV}$ space (see Figure~\ref{fig:counts_phase_sBV_grid}).
Unsurprisingly, most spectra were taken near maximum light and belong to normal-bright SNe~Ia.
The sample spectra have a median $s_{BV}$ of 0.98 and a median phase of 14~days.
Figure~\ref{fig:counts_phase_sBV_grid} then identifies the data that would provide the most improvement for the next iteration of template building: NIR spectra of fast-declining SNe in general, particularly at late phases.
These observations are naturally difficult as these SNe are rare, intrinsically faint, and fade more rapidly than normal-bright SNe.

Our approach is to model the dependence of spectroscopic properties on phase and $s_{BV}$ as a hypersurface in the phase$-s_{BV}$ space (Section~\ref{sec:GPR}).
The dearth of data in certain regions and the over-reliance on individual well-observed SNe (e.g., SN~2012fr has 25 spectra in the sample) may produce potential biases if the data represent the extremes in the respective bins.
The best way to improve this is a larger sample which fills in the phase$-s_{BV}$ space.
To characterize this potential bias, we compare the templates built with and without the 25 spectra of SN~2012fr in Section~\ref{sec:templates}.

\begin{figure*}[htb!]
\centering
\includegraphics[width=0.7\textwidth]{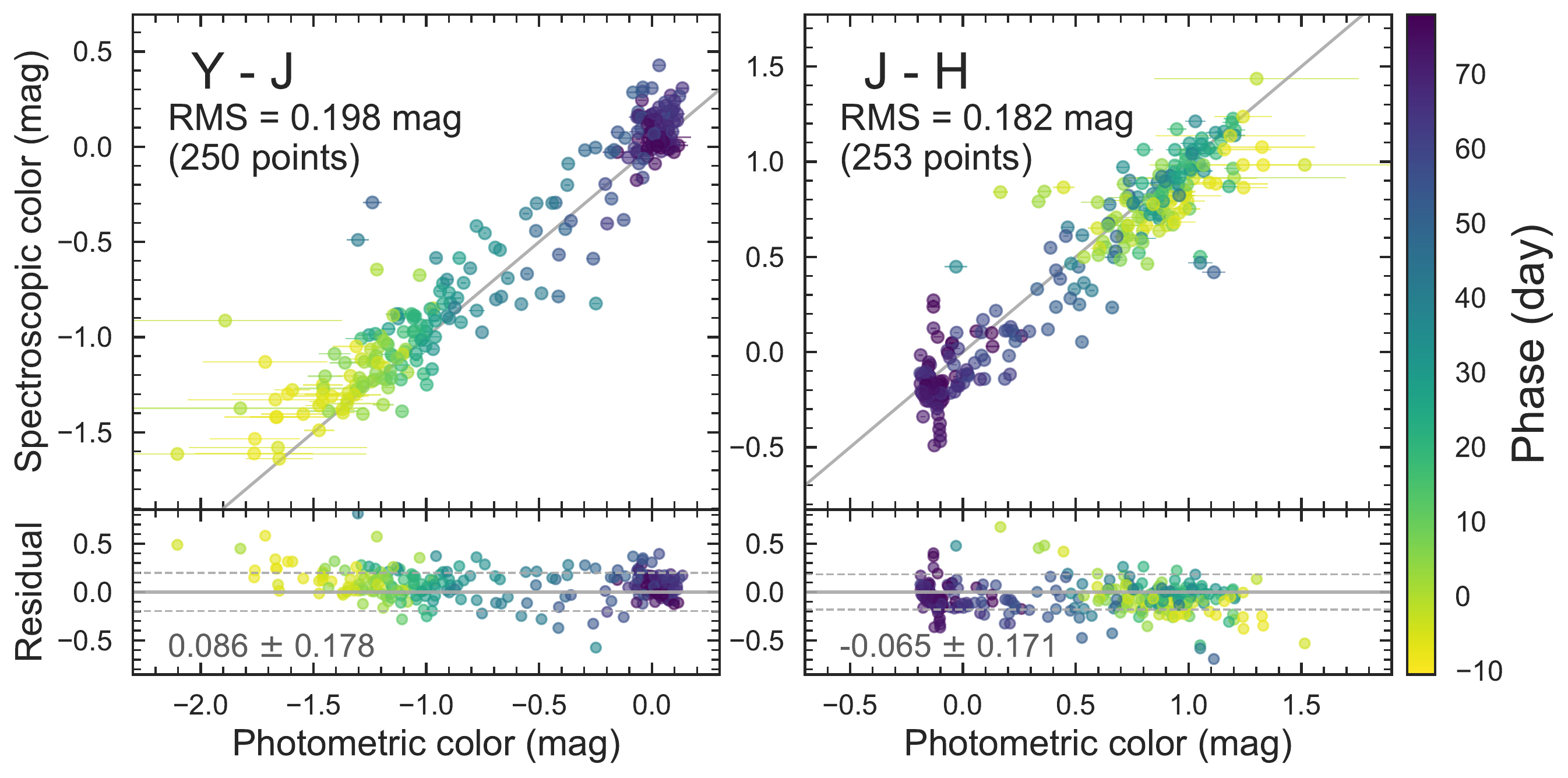} 
\caption{
Comparison between the observed spectroscopic and photometric colors for the selected sample, color-coded by phase.
Both are corrected for MW extinction.
The solid lines in the top panels correspond to the ideal one-to-one correlations. 
The resulting RMS is approximately 0.2~mag in both $Y-J$ and $J-H$, and represents the upper bound in the spectrophotometric uncertainty, as this figure includes the uncertainty in the light-curve interpolation.
In the bottom panels, the horizontal solid and dashed lines represent the mean and the dispersion of the residuals, respectively, with the values printed in the lower left corner.
}
\label{fig:spec_phot_colors}
\end{figure*}

\begin{figure}[htb!]
\centering
\includegraphics[width=0.98\columnwidth]{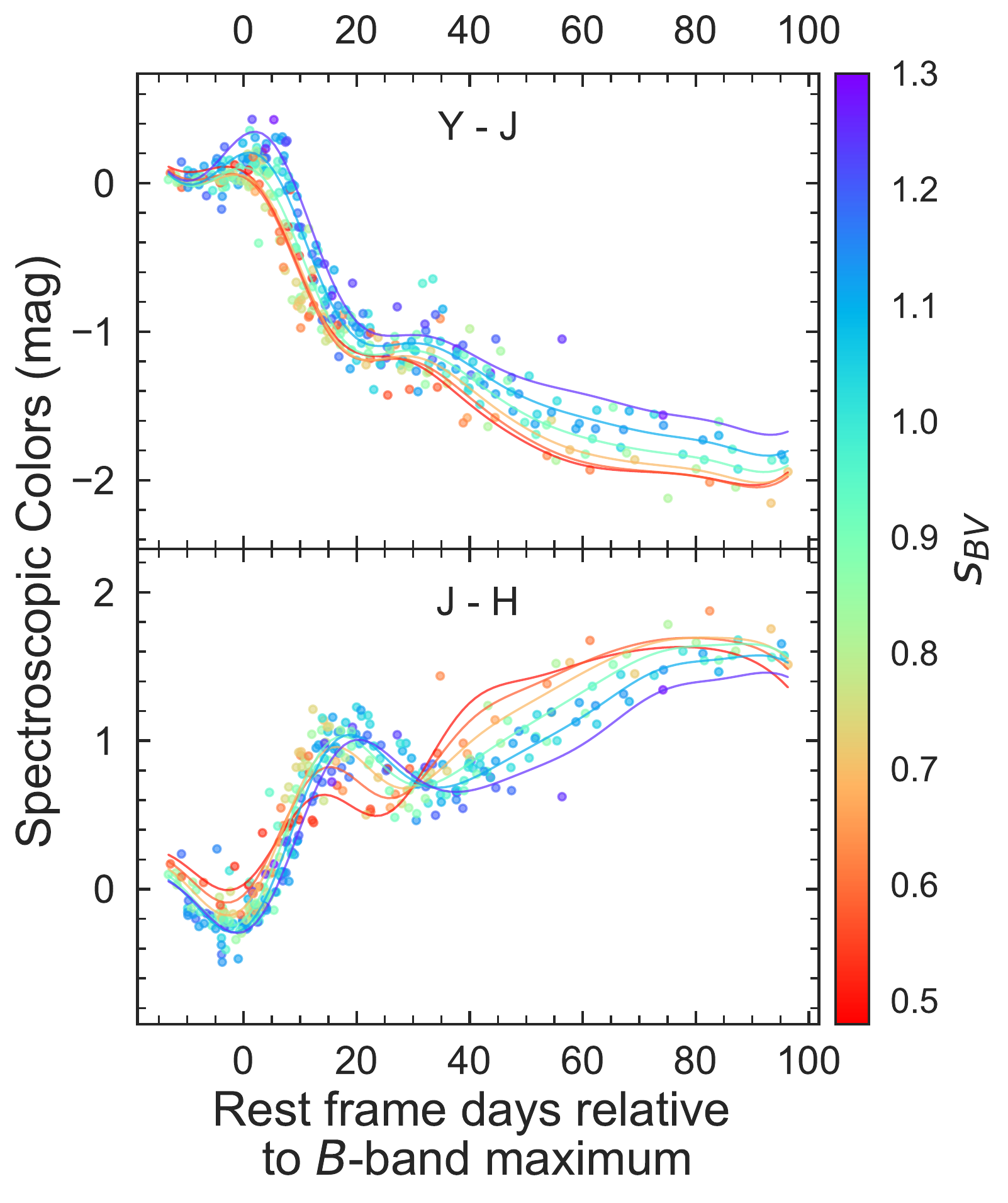} 
\caption{
Time evolution of the sample spectroscopic colors in rest frame. 
The spectra were corrected for MW extinctions.
The points are color-coded by the $s_{BV}$ of the SNe.
To aid the visualization, Gaussian process regression (GPR) fits of the phase and $s_{BV}$ dependence are plotted as solid curves.
The flux calibration of the spectra is accurate enough to preserve these correlations.
}
\label{fig:spec_YJHK_colors}
\end{figure}

\subsection{Spectrophotometric Accuracy} \label{subsec:Spectrphotometric_accruancy} 

Since there are no spectrophotometric standard stars in the NIR that can be observed by FIRE, flux calibration was done using telluric standards \citep{Hsiao2019}.
Each standard is a A0V star similar to Vega observed at a similar airmass, usually immediately before or after the science observation.
The flux calibration is performed simultaneously with the telluric correction given the $B$ and $V$ magnitudes of the A0V star, following the procedures outlined by \citet{Vacca2003}.  
Here, we examine the spectrophotometric accuracy of our sample NIR spectra calibrated using this method in terms of the relative colors.
Note that the vast majority of the spectra were observed at parallactic angle to prevent the change of spectroscopic colors caused by the slit loss.

To assess the spectrophotometric accuracy after flux calibration, the synthetic broadband $Y-J$ and $J-H$ colors measured directly from the spectra are compared to those from the corresponding light curves.
All spectra were corrected for Milky-Way (MW) extinction using the \citet{Schlafly2011} dust map and the \citet{Fitzpatrick1999} law assuming $R_V = 3.1$.
Spectroscopic colors were then measured from synthetic photometry using $YJH$ filters.
Due to the design of CSP-II, the photometric coverage has typically nightly cadence in the optical but has large gaps in the NIR \citep{Phillips2019}.
Here, we took advantage of the simultaneous optical-to-NIR light-curve fitting of \texttt{SNooPy}, using the \texttt{SNooPy} models to interpolate the sparse data in the NIR.
The MW extinction-corrected $Y-J$ and $J-H$ colors interpolated to the dates of the spectroscopic observations were then taken as the comparison photometric colors.

The comparison between the observed spectroscopic and photometric colors is shown in Figure~\ref{fig:spec_phot_colors}.
The root-mean-square (RMS) of the difference between spectroscopic and photometric colors is around 0.2~mag in both $Y-J$ and $J-H$, with the mean systematic difference within 0.1~mag. 
Furthermore, \citet{Hsiao2019} showed that the median color differences between 13 NIR spectra and well-sampled light curves are 0.08~mag in $Y-J$ and 0.03~mag in $J-H$.
This indicates that the spectrophotometric accuracy of this sample of NIR spectra (without the contribution of the interpolation uncertainty) is roughly at the $10-20$\% level.

Another way to examine the spectroscopic colors is to look at their evolution over time.
Figure~\ref{fig:spec_YJHK_colors} shows the resulting time evolution of the spectroscopic $Y-J$ and $J-H$ colors. 
There is clear evidence that the color evolution depends heavily on the $s_{BV}$ parameter, indicating that the spectroscopic colors are accurate enough to preserve such a signal.
To avoid incurring further uncertainty from the interpolation of sparse data and the narrower phase coverage of \texttt{SNooPy}, we chose \textit{not} to match the spectroscopic colors to the photometric colors in this work.


\section{Methods} \label{sec:methods}

The goal of this work is to empirically construct a set of NIR spectral templates of SNe~Ia based on observed data.
The templates aim to quantify the following spectral properties: the time evolution, the correlation with a light-curve-shape parameter, and the remaining diversity in the form of statistical flux uncertainties.
A summary of the procedure is outlined here and the details are given below.
\begin{enumerate}
\item The spectra that satisfy the selection criteria are organized onto a common rest-frame wavelength grid  (Section~\ref{sec:sample_preparation}).
\item Principal Component Analysis \citep[PCA;][]{Pearson1901} is performed in order to reduce the dimensionality of the data set and characterize the spectral properties (Section~\ref{sec:PCA}).
\item Gaussian Process Regression \citep[GPR;][]{GP_book} is then employed to model the dependence of the PCA results on phase and a light-curve-shape parameter (Section~\ref{sec:GPR}).
\end{enumerate}
The final result can be used to reconstruct the normalized SN~Ia SED, as well as the associated uncertainties, given the phase and the light-curve shape of a SN~Ia.
The focus of building the template was placed on the spectral features, since in SN~Ia cosmological studies, the broadband colors can be corrected by matching photometric colors but the spectral features cannot. 

Alternatively, we also explored a neural network approach using a conditional variational autoencoder (cVAE).
The cVAE approach is able to produce promising results similar to those from the PCA+GPR method while being more flexible, see Appendix~\ref{sec:cVAE} for a discussion of the advantages and disadvantages of this proposed method.

\subsection{Data Preparation} \label{sec:sample_preparation}

As the entire data set is observed with a single instrument and configuration by design, it is possible to create a common rest-frame wavelength grid that is representative of the pixel grid and can minimize the interpolation uncertainties.
First, an average wavelength grid is determined by taking the mean calibrated wavelength of the sample spectra at each pixel in the observer frame.
This wavelength grid was then de-redshifted by the median redshift of the sample ($z_{\rm{median}} = 0.021$).
Finally, the sampling frequency of the grid was doubled by adding a mid-point between neighboring points, forming a rest-frame wavelength grid with 3673 wavelength points. 
The resulting wavelength interval ranges from $\sim$3~\AA\ in the $Y$ band to $\sim$9~\AA\ in the $K$ band.

All sample spectra were de-redshifted to the rest frame and corrected for Milky-Way (MW) extinction using the \citet{Schlafly2011} dust map and the \citet{Fitzpatrick1999} law assuming $R_V = 3.1$.
They were then interpolated onto the common wavelength grid described above.
Note that, to prevent spurious flux values from dominating the PCA, sigma clipping was performed before the interpolation, removing all flux values 5-$\sigma$ above or below the local mean with an average window size of 100~\AA.

\begin{figure*}[bt!]
\centering
\includegraphics[width=0.94\textwidth]{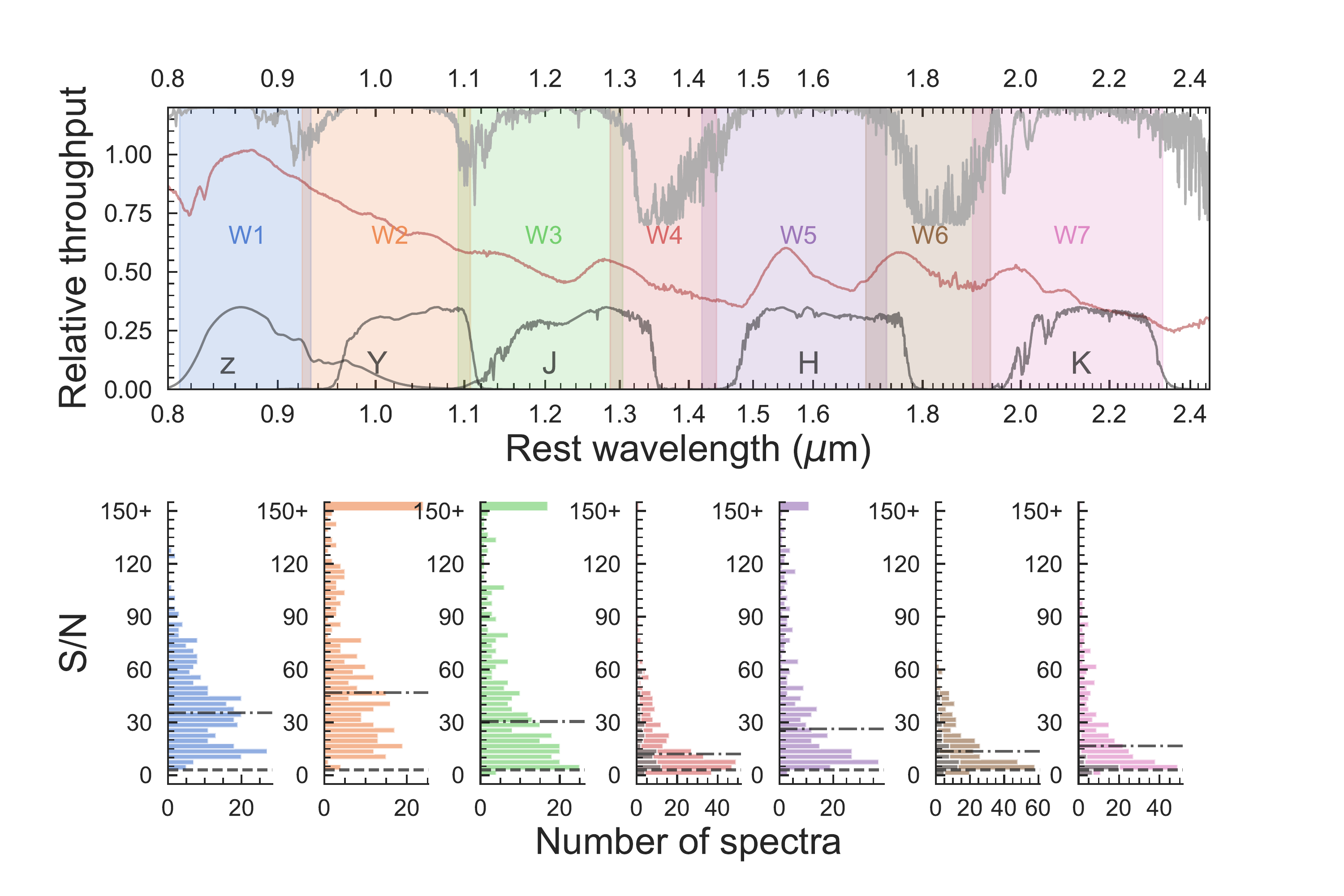}
\caption{
\textit{Top panel:} Definition of the wavelength regions. W1, W2, W3, W5, and W7 roughly correspond to the $zYJHK$ filters (dark gray solid curves), respectively, while W4 and W6 correspond to regions of strong atmosphere telluric absorptions (light gray lines on top; de-redshifted by the median redshift of the SNe sample $z=0.021$). 
A spectrum of SN~2012fr at +13~days past $B$-band maximum (red line) is plotted for illustration. 
\textit{Bottom Panels:} The distribution of the S/N of each wavelength region. 
The dash-dotted horizontal lines mark the median S/N.
The dashed horizontal lines mark the selection criterion of S/N~$\ge$~3. 
The gray stacked distributions in W4 and W6 represent the spectra that have poor telluric corrections.
The gray stacked distribution in W7 ($K$ band) indicates the spectra with spurious flux at the edge.
}
\label{fig:wave_blocks}
\end{figure*}

If PCA is performed directly on the observed spectra covering so wide of a wavelength range as the NIR, the principal components (PCs) would be dominated by regions of relatively high intrinsic flux (e.g., the $z$ and $Y$ bands) and spurious variations (e.g., the telluric regions and at the edges of the spectra).
The model SEDs need to include the more subtle variations of the spectral features.
To achieve this, dimensionality reduction was performed in seven wavelength regions.
The wavelength regions were somewhat arbitrarily defined but designed to correspond to the $zYJHK$\footnote{
$z$ as Sloan Digital Sky Survey z, $YJH$ as CSP-II du Pont $+$ RetroCam $YJH$ \citep{Phillips2019}, and $K$ as CSP-I du Pont $+$ WIRC $K_S$ \citep{Contreras2010}. The CSP filter functions are available on the web at \url{https://csp.obs.carnegiescience.edu/data/filters} and in the Python package \texttt{SNooPy} \citep{Burns2011}.
} filters with some overlap.
The overlapping region ranges from $\sim$90~\AA\ between W1\&W2 to $\sim$370~\AA\ between W6\&W7, with roughly 50 sampling points in each.
The seven wavelength regions are presented graphically in the top panel of Figure~\ref{fig:wave_blocks} and tabulated in Table~\ref{tab:W_blocks}.

Additional selection criteria were imposed on the sample spectra in each wavelength region:
\begin{itemize}
\item Each spectrum must have a median signal-to-noise ratio (S/N) $\ge$ 3.
\item In the telluric regions, W4 and W6, spectra that are over- or under-corrected for telluric absorptions were excluded by visual inspection.
\item In region W7 ($K$ band), spectra with spurious flux values at the edge of the detector were excluded by visual inspection. 
\end{itemize}
Separating a spectrum into wavelength regions has the added benefit that if the spectrum for example has poor telluric corrections, we do not have to discard other wavelength regions containing useful information.
Approximately two thirds of the sample have reliable telluric corrections, consistent with what we found in our SN~II sample using the Pa$\alpha$ feature \citep{Davis2019}.

The distributions of the S/N in each region are presented in the lower panels of Figure~\ref{fig:wave_blocks}, and the final sample count and median S/N in each wavelength region are tabulated in Table~\ref{tab:W_blocks}.
The S/N is highest in the $Y$ band and decreases toward the blue and the red.
As expected, the lowest sample count and S/N are found in the telluric regions, W4 and W6, with the median S/N staying above 10.

\begin{table}[htb!]
\centering
\addtolength{\tabcolsep}{-2pt}
\caption{Wavelength regions for dimensionality reduction.\label{tab:W_blocks}}
\hspace{-1.cm}\begin{tabular}{lcccccc}
\hline \hline
Region              & Filter\tablenotemark{a} & $\lambda_{\text{start}}$  & $\lambda_{\text{end}}$ & Median  & Sample\tablenotemark{b}   &N$_{\rm{PC}}$\tablenotemark{c}     \\
                    &                         & $(\mu\text{m})$           & $(\mu\text{m})$        & S/N     & Counts                 \\  
\hline
W1                  &  $z$                    & 0.8100                    & 0.9328                 &  35     & 331       & 11       \\
W2                  &  $Y$                    & 0.9240                    & 1.1072                 &  47     & 331       & 11       \\
W3                  &  $J$                    & 1.0925                    & 1.3041                 &  31     & 327       & 10       \\
W4\tablenotemark{d} &  $\cdots$               & 1.2865                    & 1.4423                 &  14     & 220       & 5       \\
W5                  &  $H$                    & 1.4198                    & 1.7314                 &  27     & 328       & 10       \\
W6\tablenotemark{d} &  $\cdots$               & 1.6931                    & 1.9361                 &  16     & 238       & 6       \\
W7                  &  $K$                    & 1.8989                    & 2.3300                 &  20     & 291       & 8       \\ 
\hline
\end{tabular}
\flushleft
\tablenotetext{a}{Corresponding filter overlapping the most with defined wavelength region.}
\tablenotetext{b}{Number of spectra after additional selection in each region, see Section~\ref{sec:sample_preparation}.}
\tablenotetext{c}{Number of PCs kept for template construction, see Section~\ref{sec: PC selection}.}
\tablenotetext{d}{Regions of strong  telluric absorptions.}
\end{table}

\subsection{Dimensionality Reduction} \label{sec:PCA}

A SN~Ia spectrum contains a large number of elements, but they do not vary independently.
Spectral variations are largely captured by only a few parameters \citep[e.g.,][]{Nugent1995}.
It is thus possible to drastically reduce the dimensionality of SN~Ia spectroscopic data, and PCA was chosen as the statistical technique for this purpose.
It is widely used for modeling the SN spectra \citep[e.g.,][]{Hsiao2007,Davis2019,Shahbandeh2022}.
PCA produces an orthogonal set of eigenvectors or PCs ranked by their eigenvalues, which give a measure of the amount of data variation they capture.
In this work, the matrix diagonalization was done using the Python package \texttt{scikit-learn} \citep{scikit-learn}.

As mentioned in Section~\ref{sec:sample_preparation}, the data were separated into wavelength regions in order to capture subtle variations.
In each wavelength region, flux elements were placed on a uniform wavelength grid and normalized to have the same integrated flux in the region before PCA. 
Our chosen light-curve parameter is the color-stretch parameter $s_{BV}$, since \declinerate\ and SALT \textit{x1} do not reliably classify fast decliners \citep{Burns2014}.
The PCA operation has no knowledge of the phase and $s_{BV}$ labels for each spectrum nor the wavelength value for each flux element.

\begin{figure*}
\centering
\includegraphics[width=0.97\textwidth]{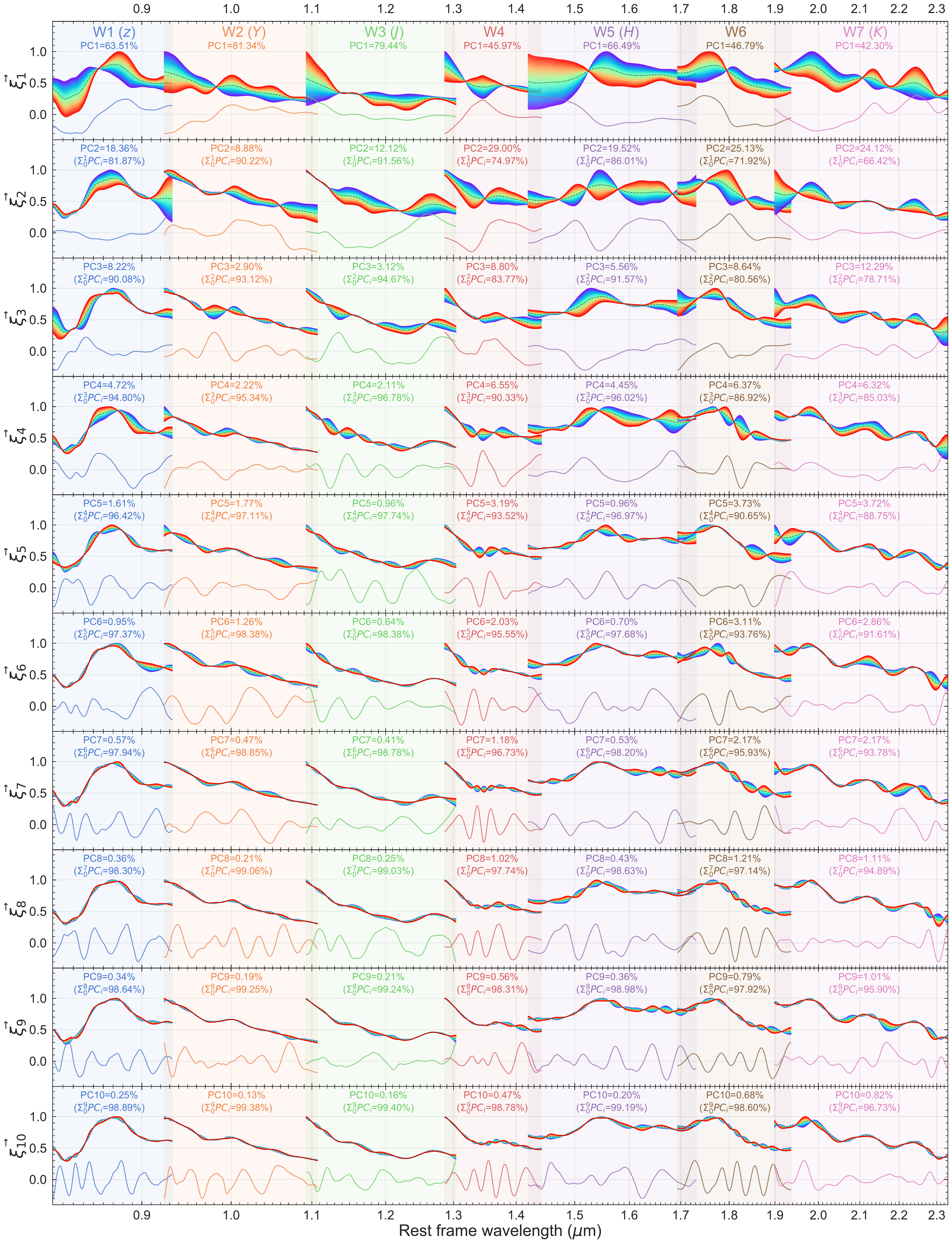} 
\caption{
The first 10 PCs and the reconstructed spectra in each wavelength region. 
The PCs are ranked by the amount of data variation they describe (labeled on top of each panel in percent variation per PC and the cumulative percent variation).
The solid curve on the bottom of each panel is the PC. 
The rainbow-colored curves are the PC-reconstructed spectra with 2$\sigma$ variation on top of the mean (dashed curve).
}
\label{fig:pca_vector}
\end{figure*}

\begin{figure*}
\centering
\includegraphics[width=0.99\textwidth]{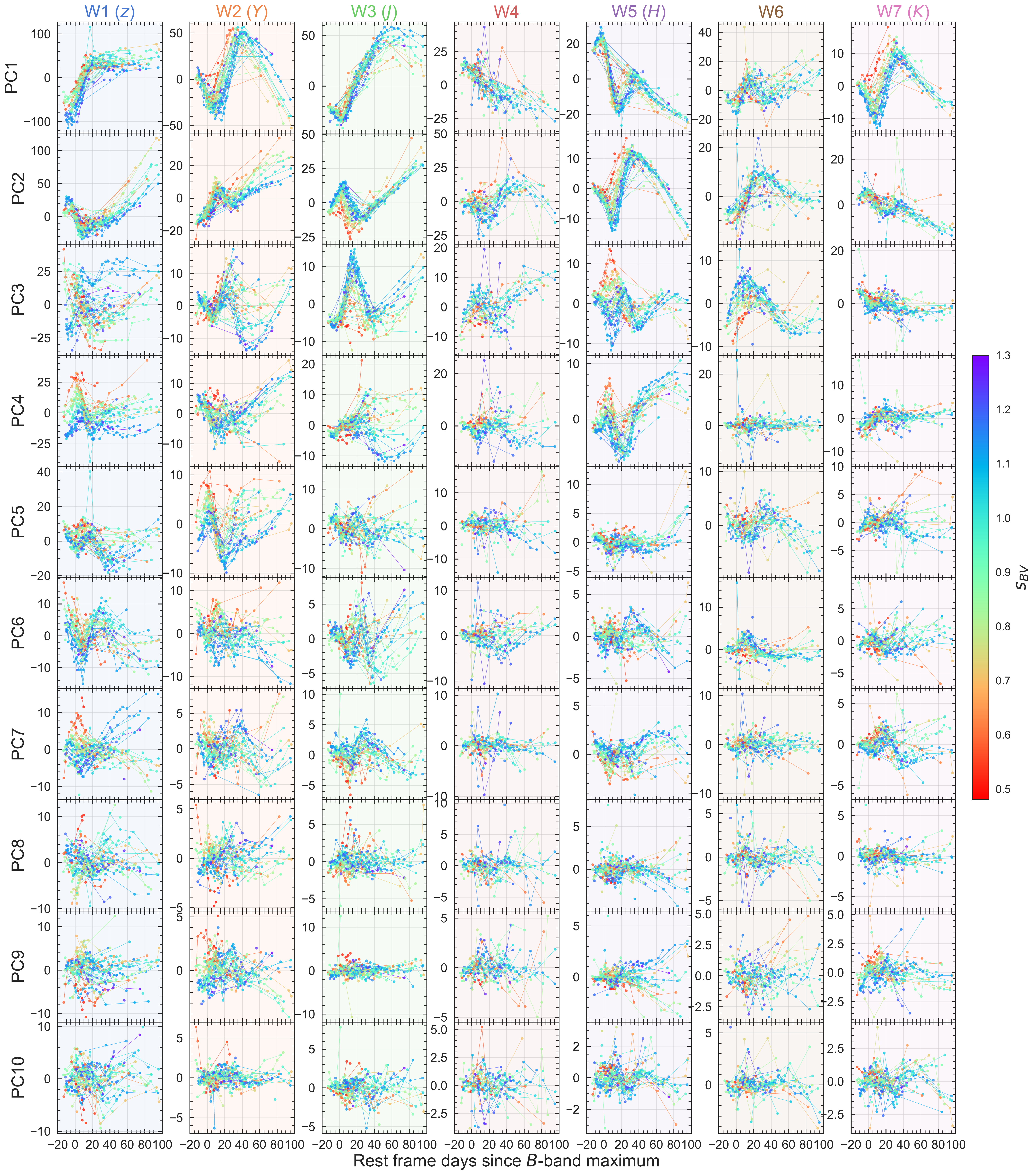} 
\caption{
The projections of the input spectra onto the first 10 PCs. 
They are plotted with respect to the phase and color-coded using their $s_{BV}$ values. 
The solid lines connect the spectra of the same SN. 
In the higher ranking PCs, the projections are highly correlated with phase and $s_{BV}$ in many PCs and wavelength regions,
for example, PC2 and PC3 in W5 ($H$ band) and PC1 and PC2 in W2 ($Y$ band).
}
\label{fig:pca_proj_phase}
\end{figure*}

The resulting PCs are presented in Figure~\ref{fig:pca_vector}.
Note that, in the figure, the PCs in the overlapped area between neighboring wavelength regions are not expected to be connected, since PCA is independently performed in each region.
In regions W1 to W6, it takes less than 5 PCs to capture $>$90\% of the variations.
Regions W2 and W3 ($Y$ and $J$ bands), in particular, only required 2 PCs each, indicating that the spectra are rather uniform.
On the other hand, W7 ($K$ band) requires 6 PCs to reach 90\% of the variation, perhaps hinting at substantial diversity.

Figure~\ref{fig:pca_vector} also shows the reconstructed spectra of 2$\sigma$ variation on top of the mean spectrum.
Familiar spectral features in the NIR then emerge, such as the \ion{Ca}{2} IR triplet in W1 ($z$ band), \ion{Mg}{2} in W2 ($Y$ band), the $H$-band break in W5 ($H$ band), and \ion{Co}{2} in W7 ($K$ band).
Another example of the captured information can be found in PC1 and PC3 of W1 and W2, which appear to describe the strength and velocity of the spectral features, respectively, in both regions.

Each input spectrum can then be projected onto a PC in the multidimensional data space, providing an associated projection value.
Mathematically, each input spectrum $\vec{f_m}$ of $n$ flux elements is then represented by the sum of all PCs $\vec{\xi_i}$ weighted by the projections $p_{i,m}$:
\begin{equation} \label{Eq:pca}
\vec{f_{m}} = \sum_{i=1}^{n} p_{i,m} \vec{\xi_i} + \vec{f_{\mathrm{mean}}}
\end{equation}
The projections are shown in Figure~\ref{fig:pca_proj_phase} with respect to phase and $s_{BV}$.
It is clear that the first several PCs in all regions have strong dependence on phase and $s_{BV}$.
The dependence is the weakest in the telluric regions W4 and W6 because of the lower S/N of the input data and the general lack of strong features.
W5 ($H$ band) shows particularly strong dependence in the top four PCs, giving confidence that the most prominent spectral features in the NIR can be well predicted with light curves alone.
This offers the motivation for the next step: modeling the dependence of spectral properties, represented by the PC projections, as a hypersurface in phase-$s_{BV}$ space.

\subsection{Modeling Parameter Dependence} \label{sec:GPR}

In the previous step, the spectral data have been reduced in dimensionality such that each spectrum is represented by a few PC projections.
Here, we will determine the dependence of the spectral properties on phase and $s_{BV}$.
The method chosen to map the projection hypersurface in phase-$s_{BV}$ space is GPR.

\begin{figure}[htb!]
\centering
\includegraphics[width=0.99\columnwidth]{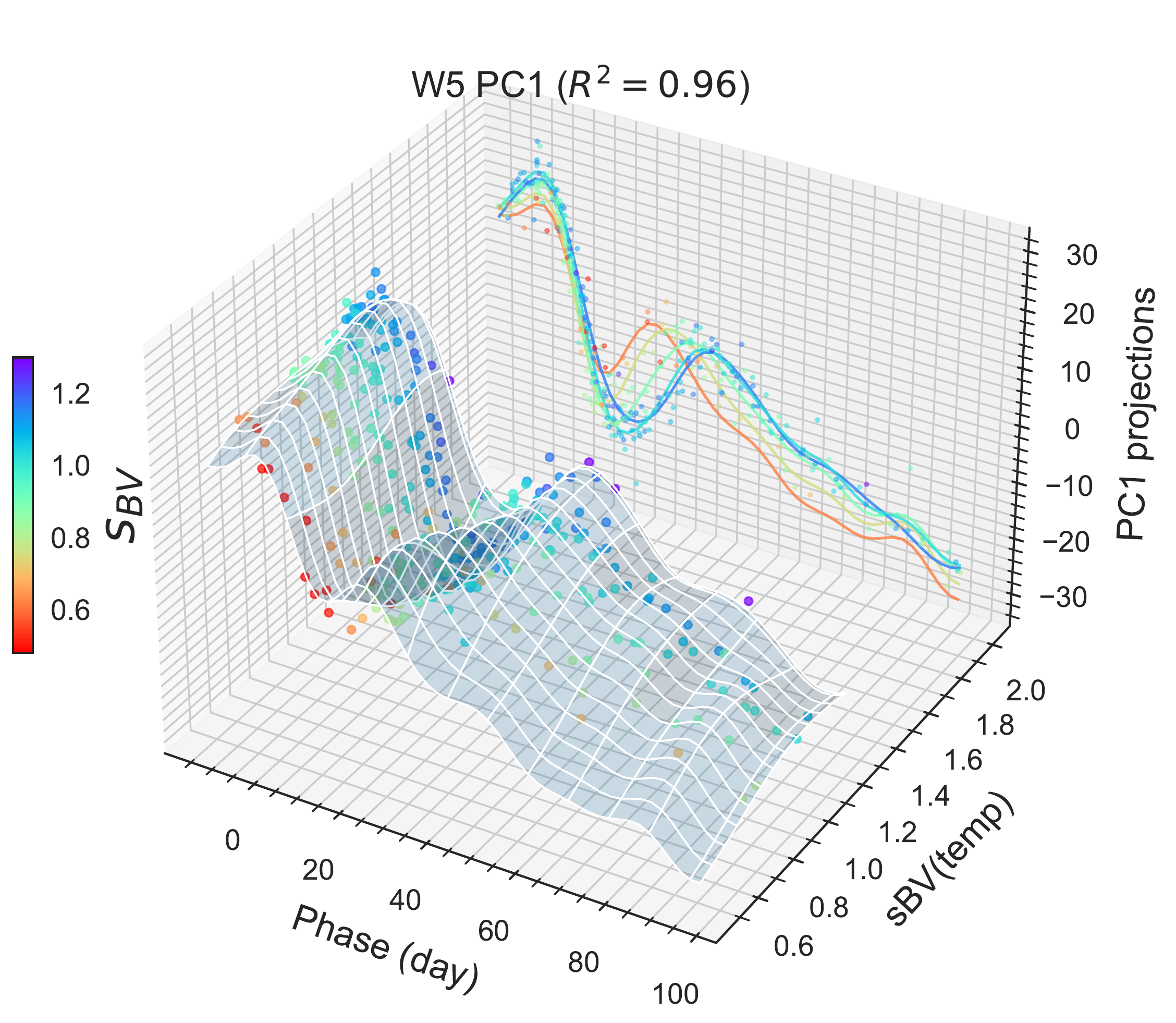} 
\caption{
An example of the GPR mapping of the projection hypersurface in phase-$s_{BV}$ space.
The side view of the hypersurface with various $s_{BV}$ slices is color coded by $s_{BV}$.
Note that the side view of the data is the same as the one plotted in Fig.~\ref{fig:pca_proj_phase}.
The $R^2$ value noted on top of the figure is the coefficient of determination of the GPR prediction.
The example shown is of PC1 in W5 ($H$ band) and has one of the highest $R^2$ values in our analysis.
}
\label{fig:GPR_example}
\end{figure}

GPR has been shown to be successful at interpolating supernova light-curve data with intermittently missing data \citep[e.g.][]{Vincenzi2019,Pessi2022}.
The data characteristics are similar here (Figure~\ref{fig:counts_phase_sBV_grid}).
It also has the advantage of producing an uncertainty estimate to the hypersurface, which allows the construction of not only the spectral templates but also the statistical flux uncertainties.
In this work, GPR was performed using the Python package \texttt{scikit-learn} \citep{scikit-learn}.
An example of the resulting hypersurface is shown in Fig.~\ref{fig:GPR_example}, successfully capturing the spectral properties as a function of phase and $s_{BV}$.

The kernel setup for the GPR in this work was a constant kernel multiplied by a radial-basis function (RBF) kernel plus a white noise kernel\footnote{The initial length scales for the RBF kernel adopted in this work are 10 and 0.1 for phase and $s_{BV}$, respectively.}.
The hyperparameters are optimized by maximizing the log-marginal likelihood. The details of the initial hyperparameter setting and optimal values can be found in Appendix~\ref{appendix:GP kernal parameters}.
Before performing GPR, the projection outliers were excluded by clipping spectra with projection values 5-$\sigma$ above or below the mean.
The remaining projection values were then normalized to range from 0 to 1 to ensure consistency across all wavelength regions and PCs while maintaining a uniform kernel setup.

With the hypersurface in hand, one can obtain the modeled PC projection value and the associated uncertainties for each PC in each wavelength region, given the phase and $s_{BV}$.
Once it was decided which PCs are to be included, template spectra were then constructed via reverse transformation to return the projections to flux space.
Finally, the spectra from individual wavelength regions were merged by matching the integrated flux in the overlapping regions and using the weighted average flux\footnote{Linear weights were assigned to the flux points such that the spectrum edges have the lowest weights.}.
Now the question becomes: what is the minimum number of PCs necessary to describe the spectroscopic diversity and how to select them efficiently? 
This is addressed in Section~\ref{sec: PC selection}.

\begin{figure*}
\centering
\includegraphics[width=0.99\textwidth]{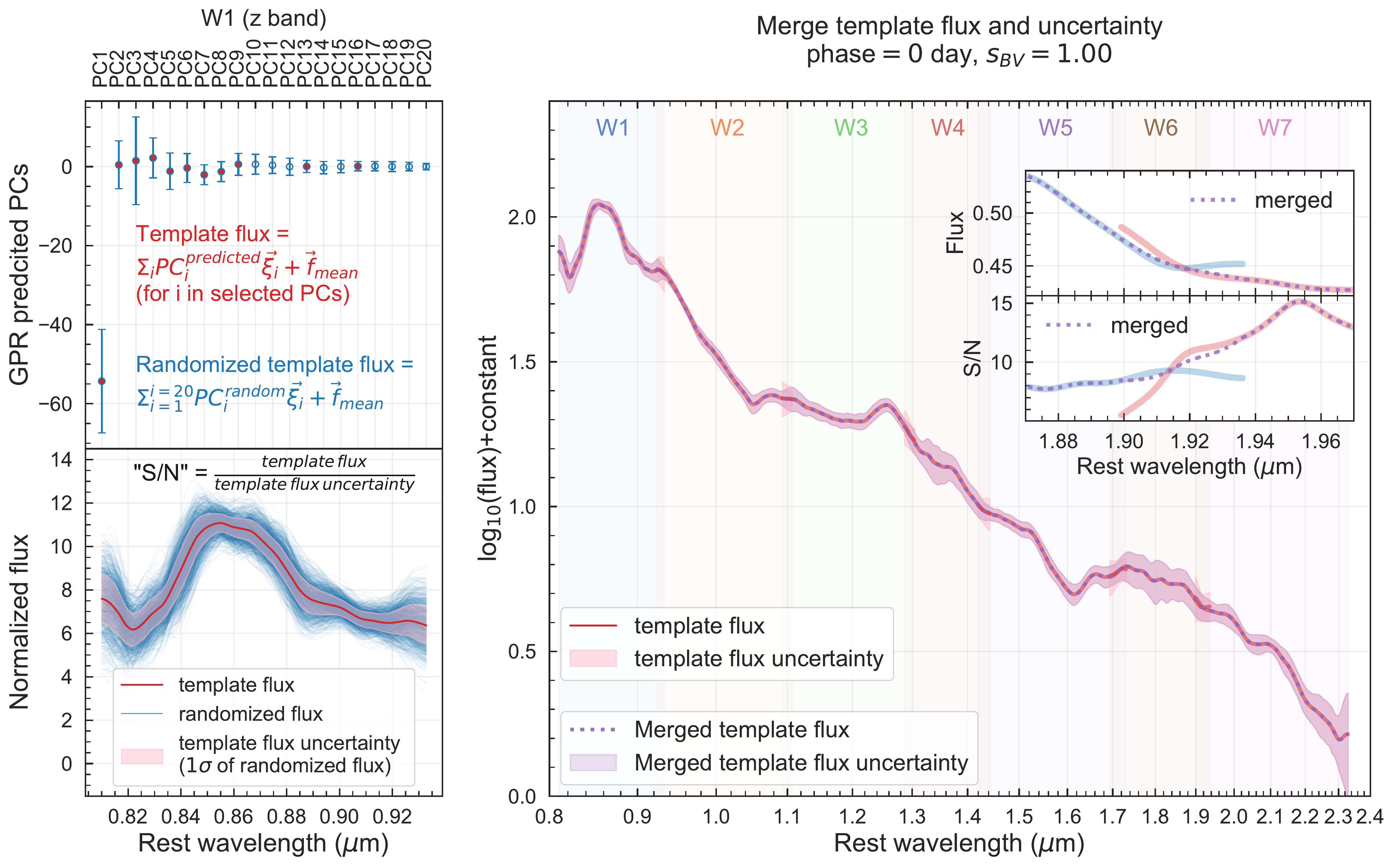} 
\caption{
Example of the template construction in W1 ($z$ band; left panel) and the merging process (right panel) of a maximum light spectrum of $s_{BV}=1$. 
In each wavelength region, the template flux is constructed by inverse-transforming the selected PCs (marked with filled red circles in the top-left example plot) that are predicted by GPR;
and the template uncertainty is represented by the standard deviation of the inverse-transformed flux of 1000 sets of randomly sampled PCs assuming Gaussian error given the GPR prediction uncertainty. 
Then the template fluxes in neighboring regions are merged together by normalizing the integrated flux and taking the weighted flux in the overlapping region.
The plotted example template in each region is normalized by the integrated flux in the overlapping area with the previous region.
The flux uncertainties are connected together through taking the weighted mean S/N in the overlapping area.
The inset in the right panel shows an example of such merging process between W6 and W7.
The division of the seven wavelength regions is plotted in the background.}
\label{fig:temp_construct_merge}
\end{figure*}

An important feature of the new template is flux uncertainty estimation, indicating the confidence of the SED prediction.
For example, the lack of data in certain parameter space translates to larger uncertainties in the template.
The flux uncertainties were determined using procedures similar to the template spectra construction and a Monte Carlo approach.
First, for each wavelength region and PC, the PC projection value was independently re-sampled 1{,}000 times, assuming that the GPR hypersurface uncertainty has a Gaussian distribution.
Crucially, we included all top 20 PCs for the uncertainty estimate, regardless of whether the PCs were included in the template construction.
This is because those PCs not included in the template construction represent the observed SED variations that were not captured by the template.
Each set of the re-sampled PC projections was then reverse transformed to the original flux space. 
The ratio of the template flux over the standard deviation of the re-sampled fluxes at each wavelength was then taken as the ``S/N''.
Finally, the S/N in the seven wavelength regions were merged by taking the weighted mean in the overlapping areas.
An example of the template construction and the merging process can be found in Figure~\ref{fig:temp_construct_merge}.
Note that the choice of the number of iterations is based on the convergence of the median S/N.

The PCA+GPR method we have presented here is robust, but has some limitations. 
First, it requires interpolations of the flux and their uncertainties onto a common wavelength grid prior to PCA. 
Secondly, PCA is a linear transformation and may require more dimensions to fully capture the variations in complex data distributions. 
Thirdly, GPR results are sensitive to the kernel design \citep[e.g.,][]{Stevance2022}.
Furthermore, the dimension reduction and template generation parts do not communicate with each other, for example, one cannot improve the other iteratively.
In Appendix~\ref{sec:cVAE}, we show that the neural network approach using cVAE can circumvent some of these drawbacks and achieve similar results.
However, the spectral features in the telluric region are not well modeled with the cVAE approach compared with the current method.

\subsection{Strategies for Selecting PCs} \label{sec: PC selection}

\begin{figure*}[htb!]
\centering
\subfigure[Percentage of total variance captured ]{\includegraphics[width=0.315\textwidth]{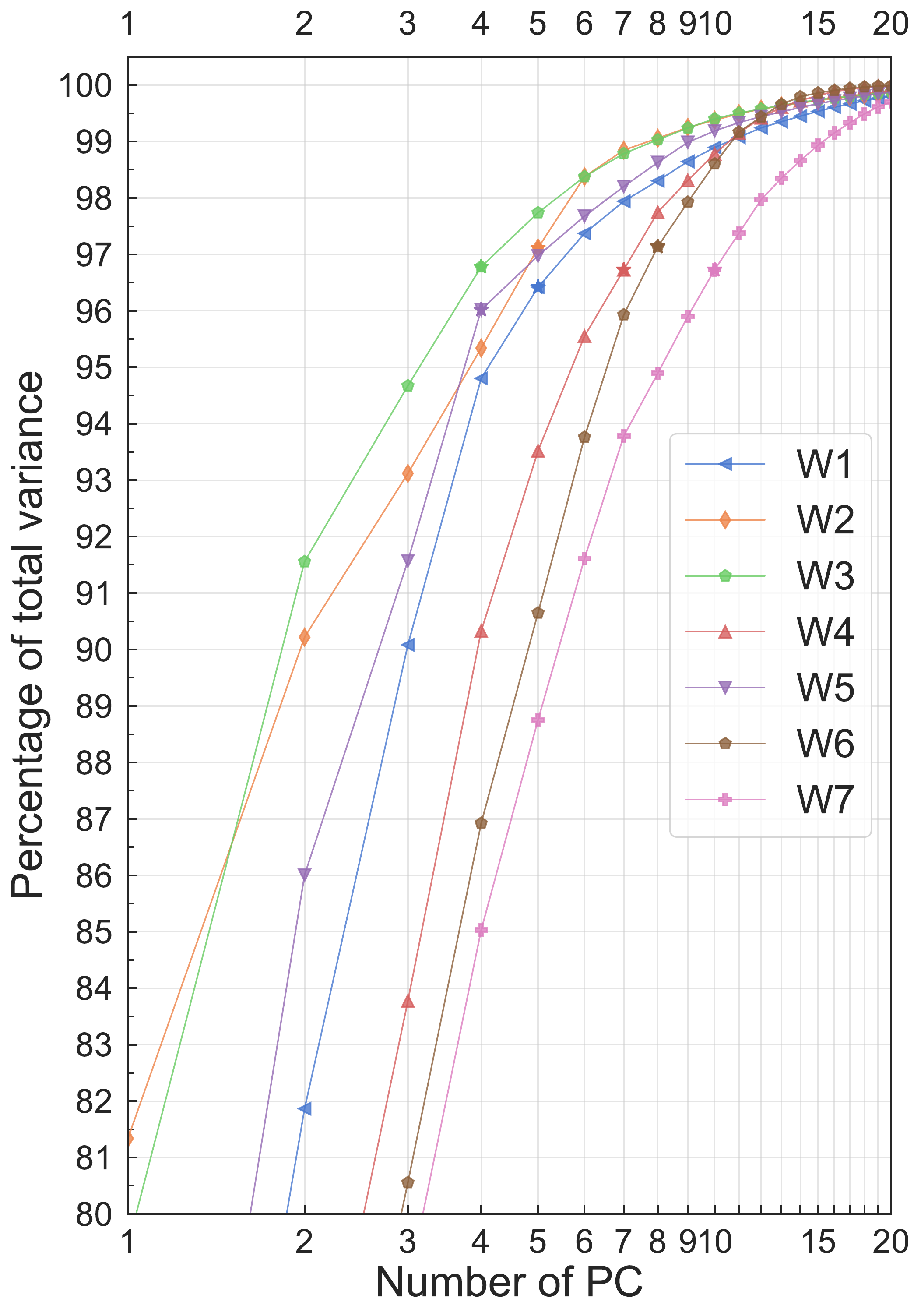}}
\hfill
\subfigure[GPR $R^2$ scores]{\includegraphics[width=0.33\textwidth]{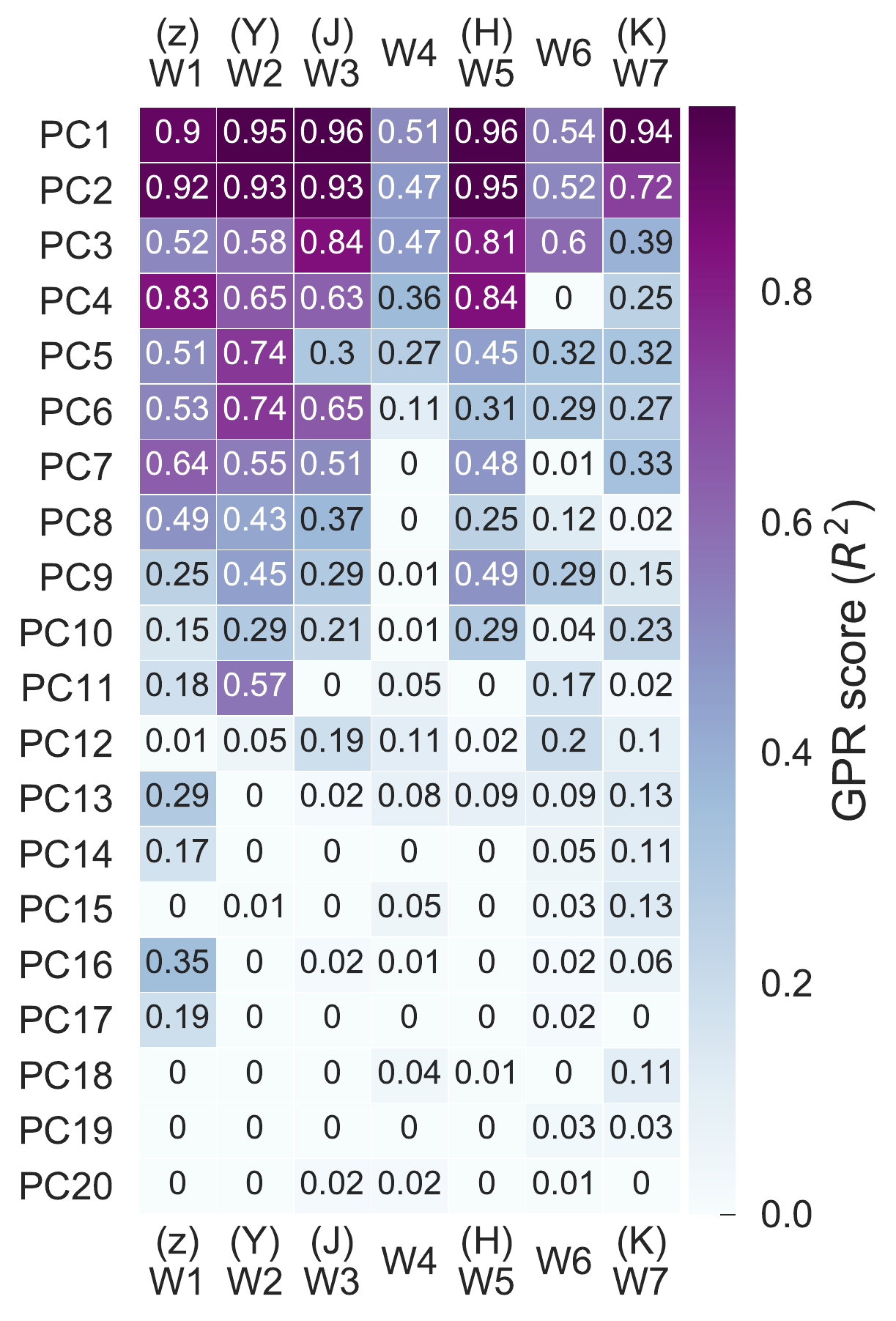}}
\hfill
\subfigure[Comparison of PC selection strategies]{\includegraphics[width=0.315\textwidth]{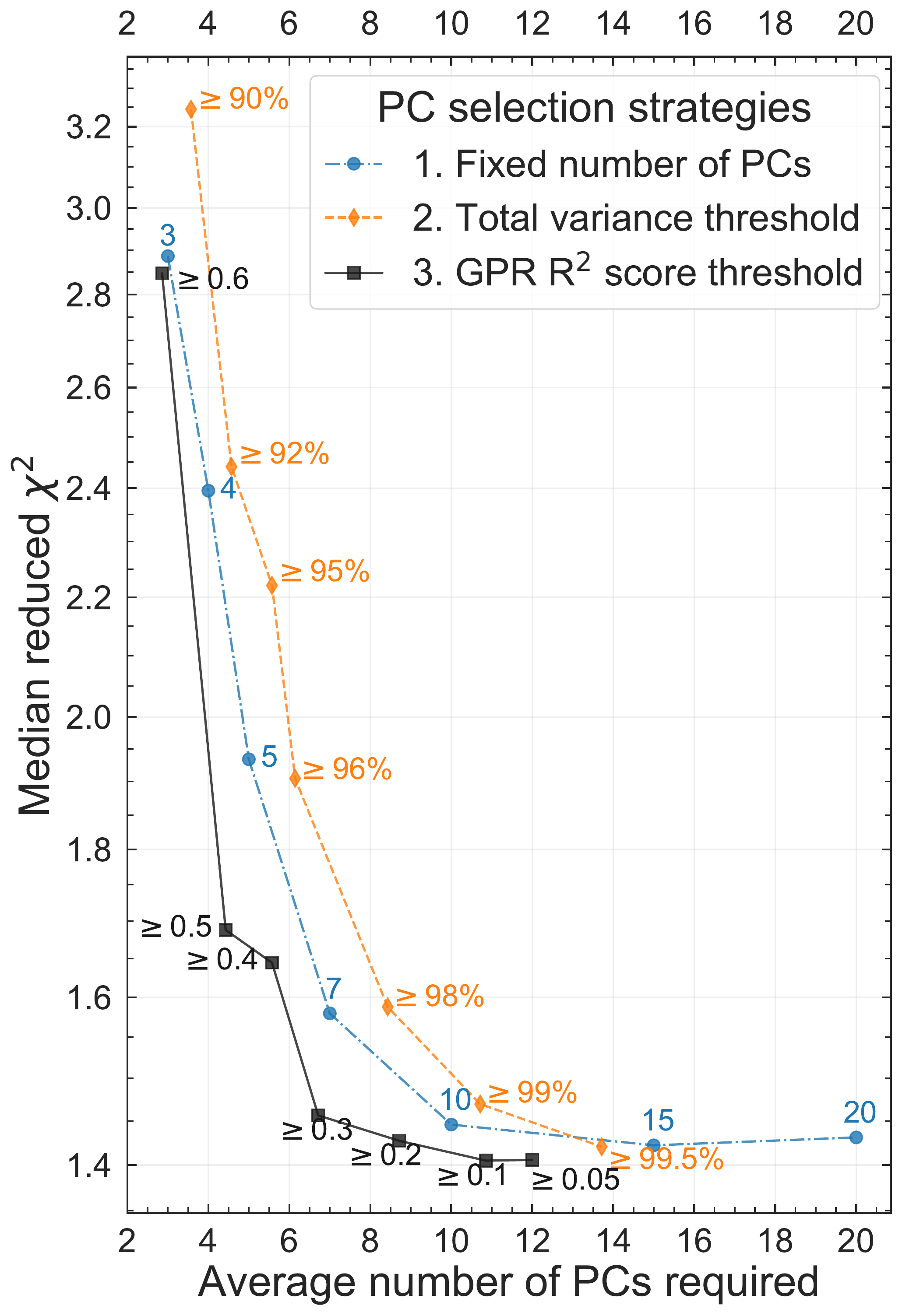}}
\caption{
\textit{(a): }The cumulative total variance percentage of the first 20 PCs in each wavelength region. 
\textit{(b): }The heat map of the GPR scores of the first 20 PCs in each wavelength region, represented by the coefficient of determination (R$^2$; which normally ranges from 0 to 1). 
Greater $R^2$ means the GPR model captures more variance in the dataset.
\textit{(c):} Comparison of three different strategies for PC selections in each wavelength region (see text for details). The y-axis is the median $\chi^2$ per degree of freedom of the spectra sample, treating the template spectra as the expected value and the (number of points -- number of PCs) as the degree of freedom. 
We adopt GPR score $R^2 \ge \GPRthreshold$ as our PC selection rule for template construction, as it efficiently reaches better $\chi^2$ per degree of freedom with fewer PCs compared to other strategies.
}
\label{fig:3panels_PCselection_strategy}
\end{figure*}

We considered three strategies for selecting the PCs for template construction:
\begin{enumerate}
    \item Use the first $N$ PCs and keep $N$ fixed for all wavelength regions.
    \item Use the first $N$ PCs that capture a total variance that exceeds a fixed threshold in each wavelength region.
    \item Use only the PCs with GPR coefficient of determination ($R^2$) scores that exceed a fixed threshold in each wavelength region.
\end{enumerate}

The $R^2$ measures how well the GPR hypersurface predicts the PC projections of sample spectra and is defined as follows: 
\begin{equation} \label{Eq:R2}
R^2 = 1 - \frac{\sum_{i}\left(p_{i}^{\mathrm{observed}}-p_{i}^{\mathrm{expected}}\right)^{2}}{\sum_{i}\left(p_{i}^{\mathrm{observed}}-\overline{p^{\mathrm{observed}}}\right)^{2}},
\end{equation}
where $p_{i}^{\mathrm{observed}}$ is the PC projection of the \textit{i}$^{th}$ spectrum;
$p_{i}^{\mathrm{expected}}$ is the GPR predicted PC projection given the same phase and $s_{BV}$ values of the \textit{i}$^{th}$ spectrum; 
and $\overline{p^{\mathrm{observed}}}$ is the mean of the PC projections of the sample spectra.
A larger $R^2$ value means that the GPR hypersurface captures more of the data variance.
A value of $R^2$ approaching 0 means that the predictions are close to the mean of projections and largely independent of the parameters.

To compare the three strategies quantitatively, we utilized the $\chi^2$ test:
\begin{equation} \label{Eq:chi2}
\chi^2 = \sum_i \frac{\left(f_i^{\mathrm{observed}} - f_i^{\mathrm{expected}}\right)^2}{\sigma_i^2} , 
\end{equation}
where $f_i^{\mathrm{observed}}$ is the observed spectrum flux of the \textit{i}$^{th}$ spectrum, $f_i^{\mathrm{expected}}$ is the predicted template spectrum flux given the same phase and $s_{BV}$ values of the \textit{i}$^{th}$ spectrum, and the observed flux errors were used as the uncertainties $\sigma_i$.
Note that the template uncertainties are not considered since they might be correlated with the observed flux uncertainties.
Each $f_i^{\mathrm{observed}}$ used for the test was color-corrected to match the color of $f_i^{\mathrm{expected}}$ before the comparison.
The degrees of freedom (dof) were taken as the number of flux points minus the number of PCs used.
Excluding the spectra with poor telluric correction and spurious flux values at the $K$-band edge (see Section~\ref{sec:sample_preparation}), 225 observed spectra among the sample were used for the $\chi^2$ test.

Each wavelength region requires a different number of PCs to reach a total variance threshold.
For example, the telluric regions W4 and W6, as well as W7 ($K$ band) require more PCs to capture 95\% of the data variance than the other regions, see panel (a) of Figure~\ref{fig:3panels_PCselection_strategy}.
As expected, the GPR $R^2$ scores are the highest for the top-ranking PCs, as shown in panel (b) of Figure~\ref{fig:3panels_PCselection_strategy}.
However, they are not ordered perfectly, revealing some lower-ranked PCs with strong dependence on phase and $s_{BV}$.
Note that the GPR $R^2$ scores are generally lower in the telluric regions W4 and W6.
While this result may be revealing real spectral behavior, it more likely points to the observational challenges in these regions.

The performance of the three proposed PC selection strategies was then assessed via the $\chi^2$ test.
In panel (c) of Figure~\ref{fig:3panels_PCselection_strategy}, we plot the average number of PCs required across wavelength regions versus the median reduced $\chi^2$ for each strategy.
Using a GPR $R^2$ score threshold (strategy 3) shows a clear advantage over the other two strategies, as it requires fewer PCs in order to reach the same accuracy in the SED prediction.
For example, to reach a reduced $\chi^2 \sim 1.43$, the average number of PCs needed is 15 for strategy 1, 13 (total variance of $\ge99.5\%$) for strategy 2, and 8 ($R^2\ge0.2$) for strategy 3. 
This trend is consistent if we take the template uncertainties into consideration for the calculation, but the reduced $\chi^2$ values are consistently lower.
Hence, we adopted strategy 3 with a threshold of $R^2\ge\GPRthreshold$. 
The number of PCs used in each region are listed in Table~\ref{tab:W_blocks}.
The GPR hyperparameters of the selected PCs are all converged to the optimal values, except the RBF length scale for $s_{BV}$ of PC1 in W4 (telluric) region, see details in Appendix~\ref{appendix:GP kernal parameters}.
Note that even though some higher-ranked PCs are omitted when constructing the template flux, it is crucial to include these when estimating the uncertainties in the spectral template.

Having PC projections that are highly correlated with phase and $s_{BV}$ would be ideal for constructing spectral templates.
However, spectral features with low GRP $R^2$ scores could aid the search for a secondary parameter for improving SN~Ia standardization or even point to interesting physics.
For example, the early-phase NIR spectra of the transitional SN~Ia iPTF13ebh has been shown to have strong \ion{C}{1} features, most notably in the $Y$ band \citep{Hsiao2015}.
These spectra also have the most extreme PC projection values compared with other spectra at early phases, such as PC2, PC6, PC7, PC8 and PC10 in W2 ($Y$ band; see Figure~\ref{fig:pca_proj_phase}).


\section{Results} \label{sec:results}

Following the methodology set forth in Section~\ref{sec:methods}, the NIR spectral templates of SNe~Ia and corresponding statistical uncertainties can be constructed as a function of phase and the light-curve parameter $s_{BV}$ using PCA+GPR.
In this section, we present the resulting NIR spectral templates and check how well they reproduce observed spectra.
Here, we use K-corrections to assess the accuracies of the templates, even though they do not need to be explicitly calculated for spectrophotometric experiments or SED-based light-curve fitters.

The new NIR spectral template is available on the CSP website\footnote{\url{https://csp.obs.carnegiescience.edu/data}} and will be implemented in \texttt{SNooPy} \citep{Burns2011}, in where the NIR template can be attached to the optical region of the time-stretched ($t_{\mathrm{stretched}} = t/s_{BV}$) Hsiao template \citep{Hsiao2007}.

\subsection{Spectral Templates} \label{sec:templates}
\begin{figure}
\centering
\includegraphics[width=0.99\columnwidth]{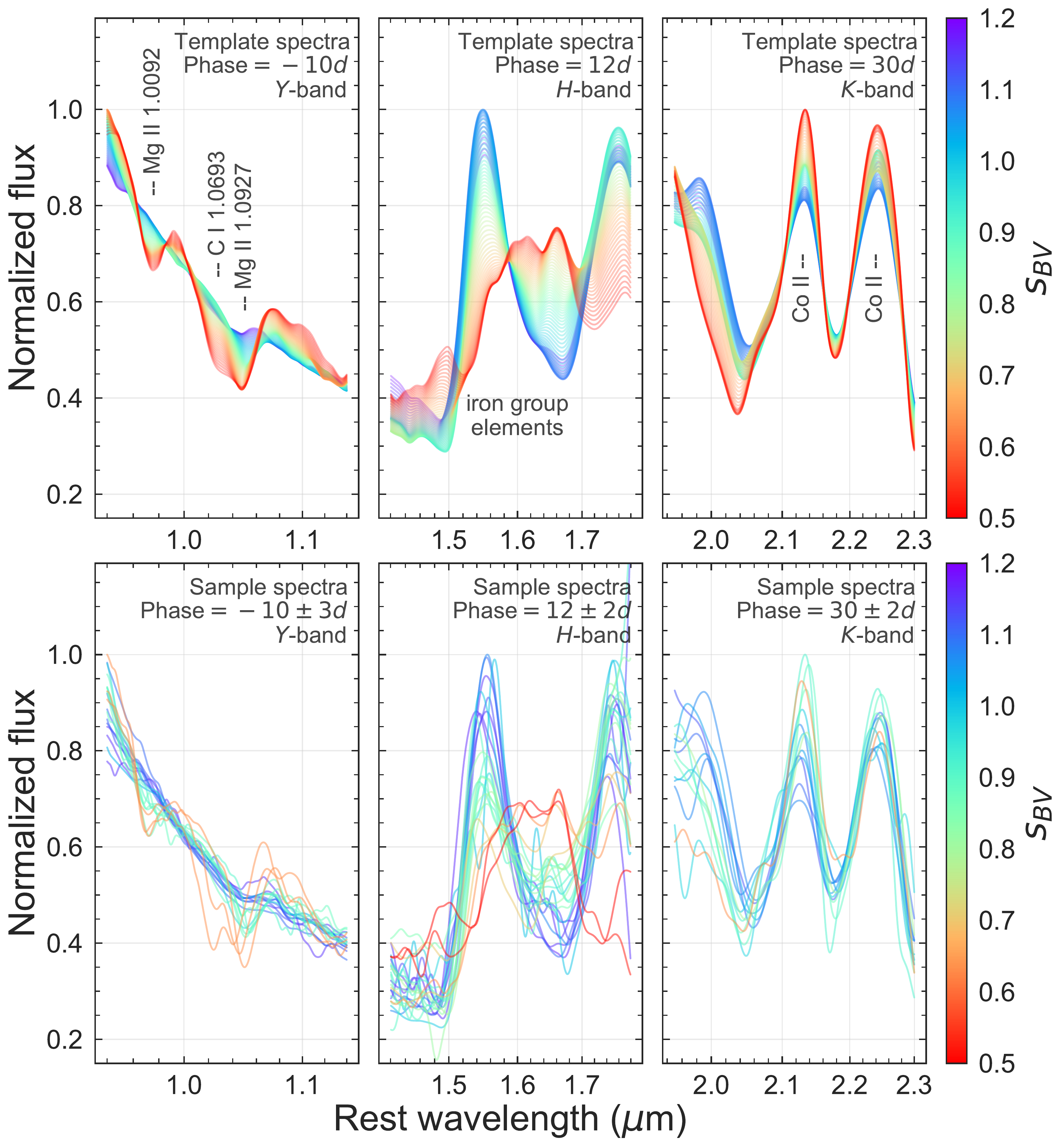}
\caption{
Comparisons between template (top panels) and observed spectra (bottom panels) for a range of light-curve color-stretch parameter $s_{BV}$.
The columns present the diversity for pre-maximum spectra around $-10$~days in the $Y$ band (left), post-maximum spectra around $12$~days in the $H$ band (center), and post-maximum spectra around $30$~days in the $K$ band.
In each panel, the spectra are normalized to have the same integrated flux in the wavelength region shown.
}
\label{fig:zoom_in_temp_obs}
\end{figure}

The template spectra constructed using the PCA + GPR method are able to replicate the spectral diversity in the sample spectra. 
In Figure~\ref{fig:zoom_in_temp_obs}, we highlight the spectral dependence on $s_{BV}$ in three wavelength regions and epochs: pre-maximum in the $Y$ band, and post-maximum in the $H$ and $K$ bands.

\begin{figure}
\centering
\includegraphics[width=0.98\columnwidth]{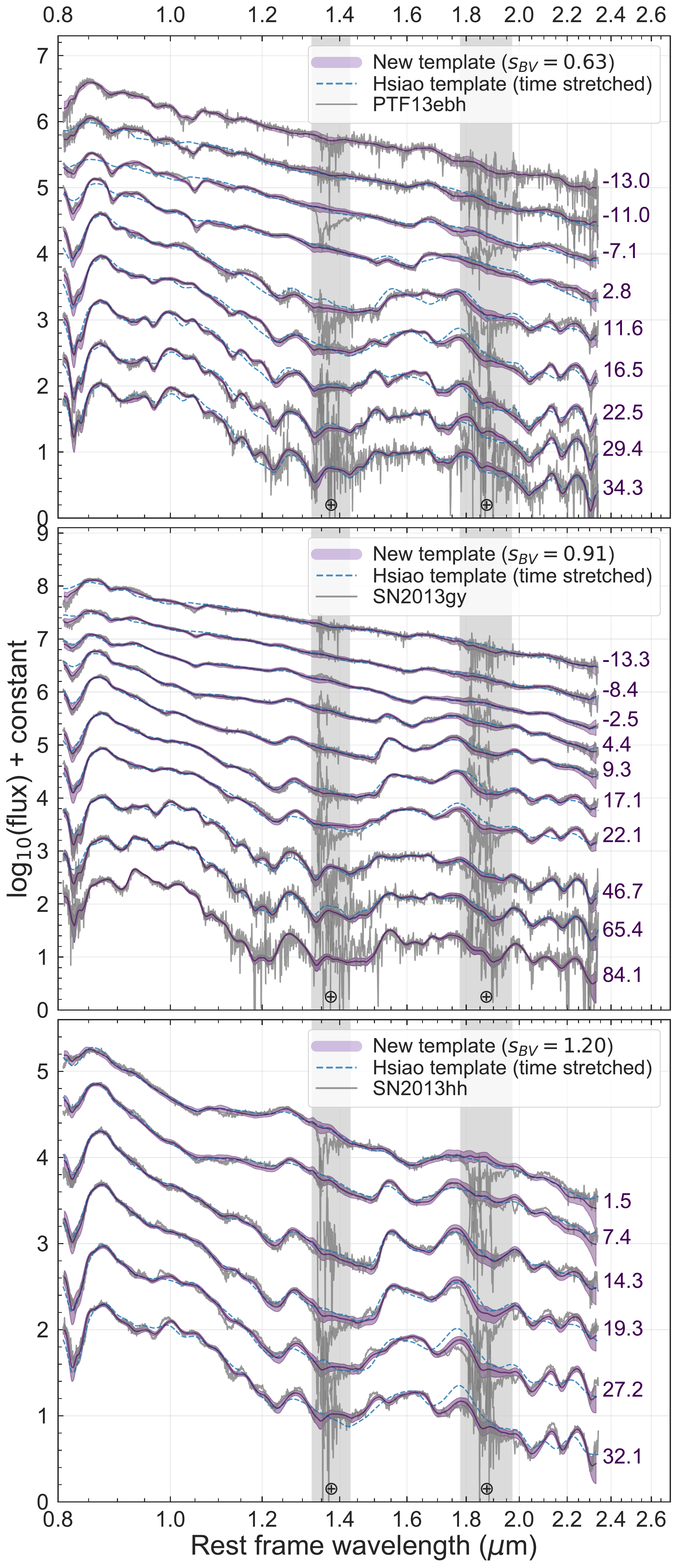}
\caption{
Comparison between template and observed spectra in the training sample.
The panels show three well-observed time series from the FIRE sample: (top) $s_{BV}=0.63$, a fast decliner, (middle) $s_{BV}=0.91$, an average object, and (bottom) $s_{BV}=1.20$, a slow decliner.
The new spectral template is shown by the solid colored curves with its 1-$\sigma$ uncertainty, while the Hsiao template is shown as blue dashed curves.
All observed spectra were color corrected to match the colors of the template spectra for presentation.
}
\label{fig:temp_comp_obs_FIRE}
\end{figure}

The left panels of Figure~\ref{fig:zoom_in_temp_obs} present the pre-maximum spectra in the $Y$ band at around $-10$~days.
At these early phases, the photosphere arises in the outer and intermediate layers of the ejecta.
Faster declining and fainter SNe clearly show stronger \ion{Mg}{2} $\lambda$~1.0092 and $\lambda$1.0927~\um\ features.
The additional absorption feature on the blue wing of \ion{Mg}{2} $\lambda$1.0927~\um\ is captured by the template and appears to be only present in subluminous SNe~Ia.
This feature has been identified to be unburned \ion{C}{1} $\lambda$1.0693~\um\ in several fast decliners: SN~1999by \citep{Hoeflich2002}, iPTF13ebh \citep{Hsiao2015}, SN~2012ij \citep{Li2022}, and SN~2015bp \citep{Wyatt2021}.
An alternative identification of \ion{He}{1} $\lambda$1.0830~\um\ has also been proposed in the context of the double-detonation scenario \citep{Boyle2017}.

At around two weeks past maximum, the prominent $H$-band break is at its peak strength \citep[e.g.,][]{Hsiao2013}.
The template shows drastic differences in the profile shapes between slow and fast declining SNe~Ia (middle panels of Figure~\ref{fig:zoom_in_temp_obs}).
These variations have been investigated in previous studies.
The strength of the $H$-band break at its peak varies strongly with the light-curve decline rate \citep{Hsiao2013,Hsiao2015}.
The velocity of the blue edge of the $H$-band break measures the $^{56}$Ni distribution and is also strongly correlated with $s_{BV}$ \citep{Ashall2019a,Ashall2019b}.
In the region between $1.6-1.7$~\um, the spectral local minimum in a slow decliner is in contrast to a local emission feature in a fast decliner, which may be the result of a photosphere well within the central Ni-rich region \citep{Hoeflich2002}.
Any viable explosion scenario needs to be able to explain these trends.

At approximately one month past maximum, the photosphere has receded to reveal the iron-peak elements in the inner layers.
The $K$ band is dominated by multiple lines of \ion{Co}{2} at this epoch \citep[e.g.,][]{Gall2012}.
Interestingly, the template only shows a mild dependence of the feature shapes on the light-curve shape (right panels of Figure~\ref{fig:zoom_in_temp_obs}).
The observed spectra show a range of feature strength but a weaker correlation with $s_{BV}$ compared to the $H$ band.
This perhaps points to a secondary parameter or contribution from a different spectral line that is not associated with the decay products of $^{56}Ni$.

In Figure~\ref{fig:temp_comp_obs_FIRE}, we present the time-series comparison between our new template, the template of \citet{Hsiao2007} updated to include the NIR \citep{Hsiao2009}, and the observed spectra within our sample.
Note that the Hsiao template gives the average SED that is designed to match a stretch $s=1$ SN~Ia.
For the Hsiao template, the variation with light-curve shape is accounted for through matching the time on the stretched time axis ($t_{\mathrm{stretched}} = t/s_{BV}$), assuming that a fast decliner also has a more rapid spectral evolution.
Within the training sample, three well-observed SNe were selected for comparison: iPTF13ebh ($s_{BV}=0.63$) representing a fast decliner, SN~2013gy ($s_{BV}= 0.91$) representing an average object,  SN~2013hh ($s_{BV}=1.20$) representing a slow decliner.
As expected, the Hsiao template performs well when tasked to match an average SN~Ia.
On the other hand, the new template performs well for all $s_{BV}$ and shows a clear advantage at the extremes.
This result demonstrates that the spectral feature dependence on the light-curve shape cannot be solely described by the speed of the evolution.

\begin{figure*}
\centering
\subfigure{\includegraphics[width=0.499\textwidth]{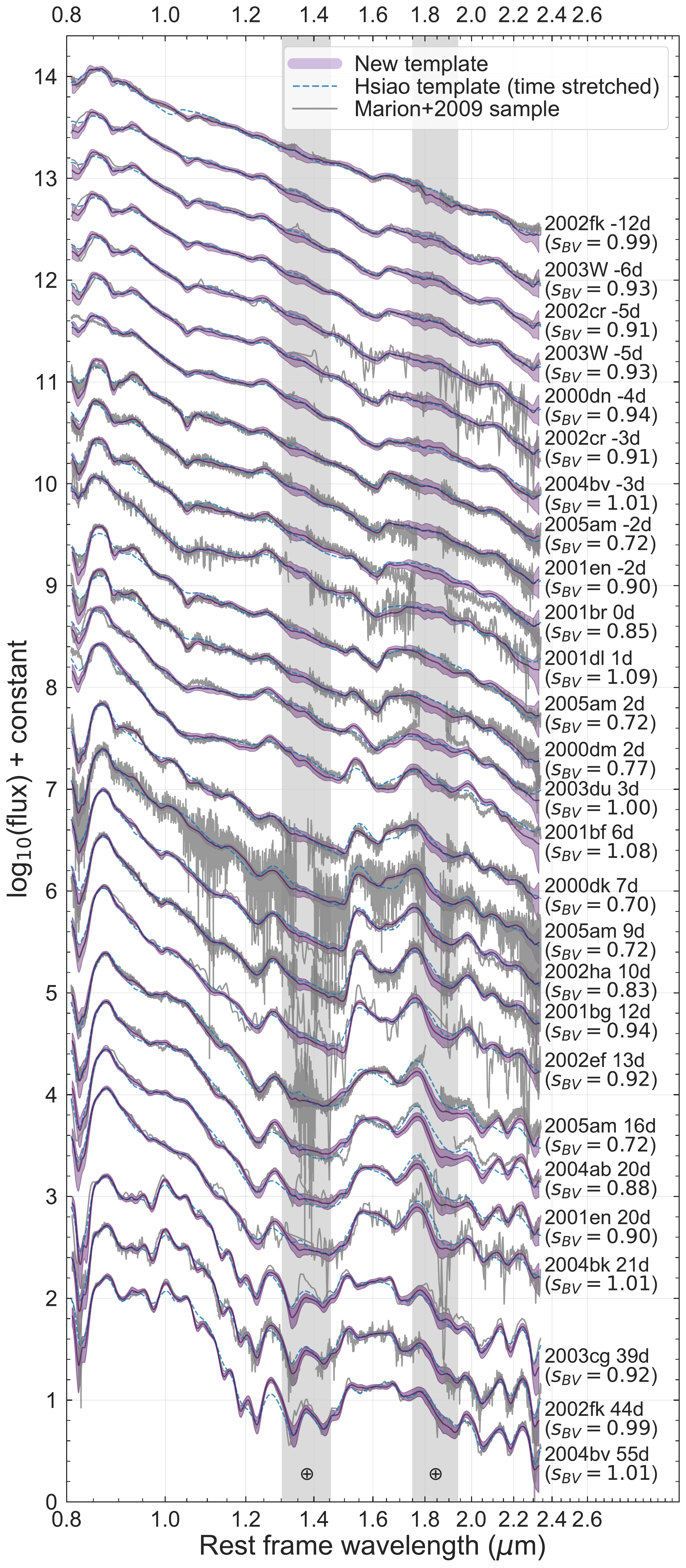}}
\subfigure{\includegraphics[width=0.49\textwidth]{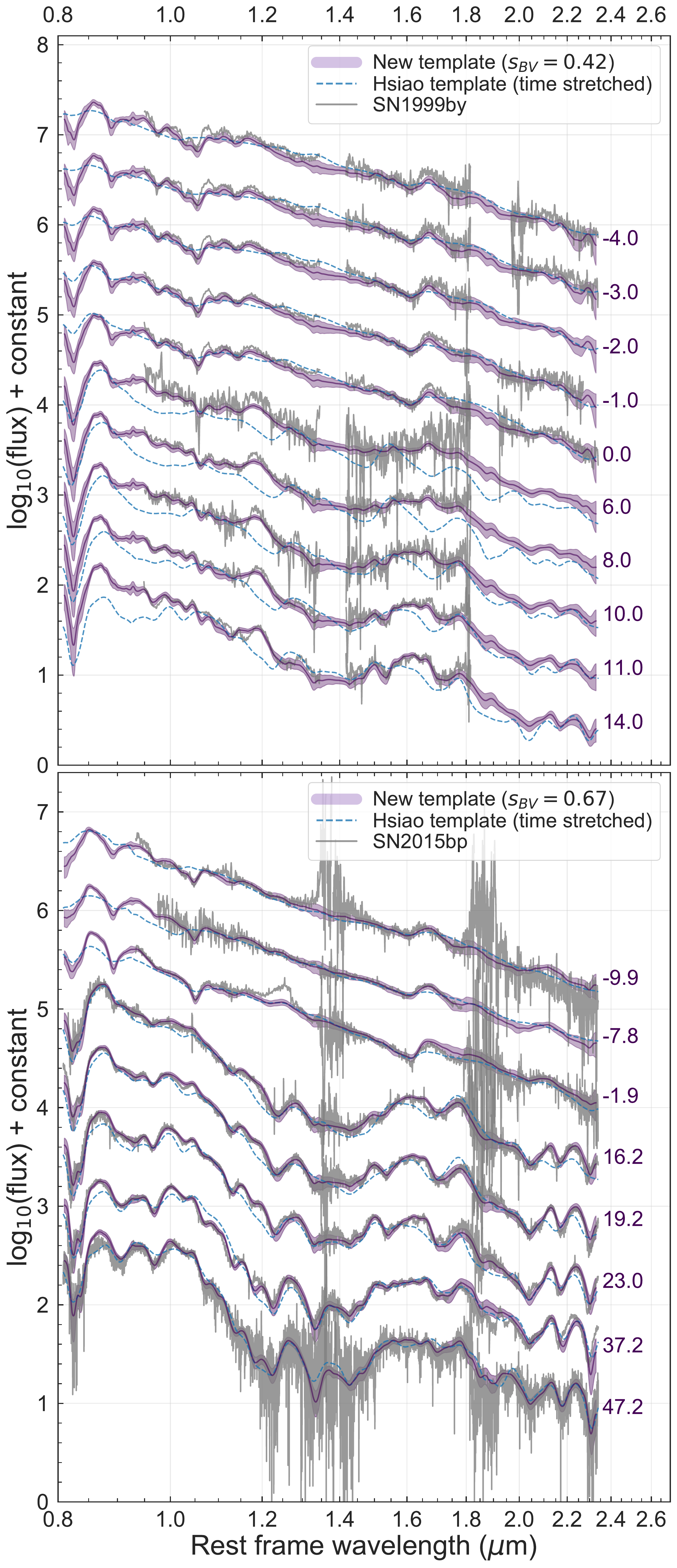}}
\caption{
Comparison between template and observed spectra that were \textit{not} included in the training sample.
The new spectral template is shown by the solid colored curves with its 1-$\sigma$ uncertainty, while the Hsiao template is shown as blue dashed curves.
The left panel shows comparisons with the sample of \citet{Marion2009}, and the template is color-coded by $s_{BV}$.
The right panels show two subluminous SNe~Ia from literature: (top) SN~1999by \citep{Hoeflich2002} with $s_{BV}=0.63$,  and (bottom) SN~2015bp \citep{Wyatt2021} with $s_{BV}=0.67$.}
All observed spectra were color corrected to match the colors of the template spectra to focus on the spectral features.
\label{fig:temp_comp_obs_non_FIRE}
\end{figure*}

\begin{table}
\setlength{\tabcolsep}{3pt}
\caption{List of SNe~Ia used for spectral comparison from the sample of \citet{Marion2009}.} \label{tab:Marion09_SNe}
\centering
\begin{tabular}{lcccc}
\hline\hline
SN      &  $z$ &\declinerate   & \declinerate\ ref.     &   $s_{BV}$\tablenotemark{a}  \\
        &    & (mag)         &                        &     \\
\hline   
2000dk  &  0.017   & 1.66           & \citet{Wang2019}        & 0.70  \\
2005am  &  0.008   & 1.61           & \citet{Hoeflich2010}    & 0.72  \\
2000dm  &  0.015   & 1.51           & \citet{Wang2019}        & 0.77  \\
2002ha  &  0.014   & 1.37           & \citet{Wang2019}        & 0.83  \\
2001br  &  0.021   & 1.32           & \citet{Wang2019}        & 0.85  \\
2004ab  &  0.006   & 1.27           & \citet{Chakradhari2018} & 0.88  \\
2001en  &  0.016   & 1.23           & \citet{Wang2019}        & 0.90  \\
2002cr  &  0.010   & 1.19           & \citet{Wang2019}        & 0.91  \\
2002ef  &  0.024   & 1.18           & \citet{Wang2019}        & 0.92  \\
2003cg  &  0.004   & 1.17           & \citet{Wang2019}        & 0.92  \\
2003W   &  0.020   & 1.15           & \citet{Wang2019}        & 0.93  \\
2001bg  &  0.007   & 1.14           & \citet{Wang2019}        & 0.94  \\
2000dn  &  0.032   & 1.13           & \citet{Wang2019}        & 0.94  \\
2002fk  &  0.007   & 1.02           & \citet{Wang2019}        & 0.99  \\
2003du  &  0.017   & 1.00           & \citet{Wang2019}        & 1.00  \\
2004bk  &  0.023   & 0.98           & \citet{Wang2019}        & 1.01  \\
2004bv  &  0.011   & 0.98           & \citet{Wang2019}        & 1.01  \\
2001fe  &  0.014   & 0.96           & \citet{Wang2019}        & 1.02  \\
2001bf  &  0.016   & 0.83           & \citet{Wang2019}        & 1.08  \\
2001dl  &  0.021   & 0.80           & \citet{Marion2003}      & 1.09  \\
\hline
\end{tabular}
\tablenotetext{a}{Converted from \declinerate\ using equation (4) of \citet{Burns2014}.}
\end{table}

Ideally, the performance of the new template should be assessed using an independent sample.
In the left panel of Figure~\ref{fig:temp_comp_obs_non_FIRE}, the new and Hsiao templates are compared to the sample of \citet{Marion2009}.
Note that the phases relative to $V_{max}$ listed by \citet{Marion2009} were converted to phases relative to $B_{max}$ assuming that $V_{max}$ occurs 2 days after $B_{max}$.
Also note that \declinerate\ values were converted to $s_{BV}$ using Equation~4 of \citet{Burns2014}.
The sources of \declinerate\ values are tabulated in Table~\ref{tab:Marion09_SNe}.
Overall, the new template provides a better match than the Hsiao template especially for low and high $s_{BV}$ values, such as SN~2005am ($s_{BV} = 0.72$).
There are three notable disagreements for the NIR \ion{Ca}{2} triplet: SN~2003W at $-$6~days, SN~2004bv at $-$3~days, and SN~2001bf at 6~days.
These could point to real diversity in the feature that is not captured by the template or observational artifacts near the short wavelength limit of the detector.


As discussed in Section~\ref{subsec:sample_distribution}, the small number of NIR spectra of subluminous SNe~Ia, especially at the late phase, is a limiting factor for the template at this particular phase space. 
Here, the template is compared to the spectra of two subluminous SNe from the literature: SN~1999by ($s_{BV} = 0.42$; \citealt{Hoeflich2002}) and SN~2015bp ($s_{BV} = 0.67$; \citealt{Wyatt2021}), shown in the right panel of Figure~\ref{fig:temp_comp_obs_non_FIRE}.
These spectra were not included in our training sample.
The spectra of SN~2015bp at various phases, including at more than 1 month past maximum, are well represented by the new template. 
This is despite the time gap of 50~days in the four FIRE spectra of SN~2015bp included in the training sample.
Because of the limited range in the light-curve shape of our sample, reproducing the spectral features of SN~1999by with such a low $s_{BV}$ requires extrapolation on the hypersurface with GPR.
Nevertheless, the spectral features of the template match exceptionally well to those of SN~1999by.
Note especially that the \ion{C}{1} features at early times were reproduced and are stronger than those of SN~2015bp.
The Hsiao template simply fails at these low $s_{BV}$ values.


\begin{figure}[htb!]
\centering
\includegraphics[width=0.85\columnwidth]{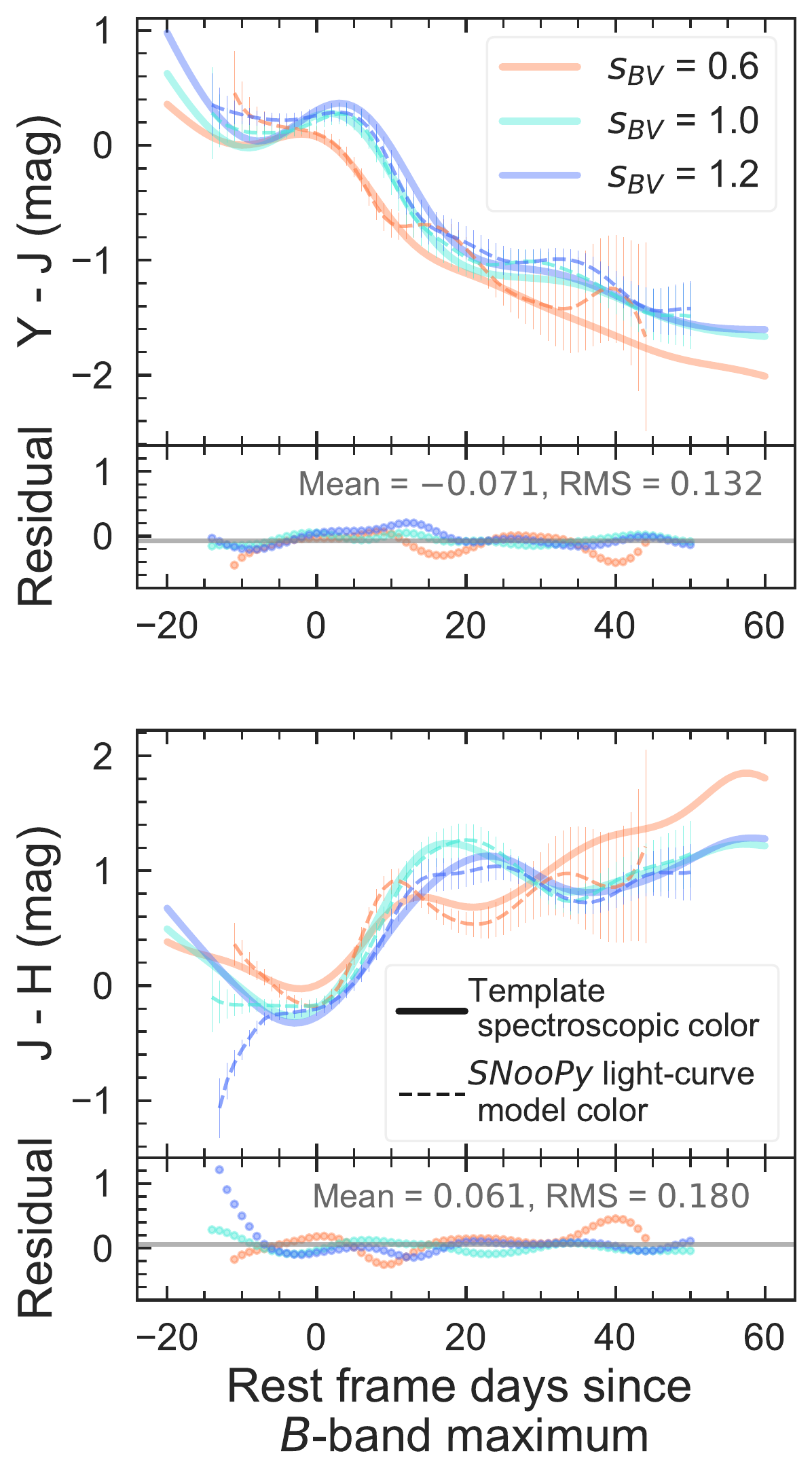} 
\caption{
Comparison of the template spectroscopic colors (solid curves) and \texttt{SNooPy} light-curve model colors for various $s_{BV}$ values (dashed curves).
The uncertainties on the \texttt{SNooPy} colors come from the uncertainties on the light-curve templates.
In the bottom panels, the residuals are shown along with the mean (solid gray lines).
The mean and RMS values are also labeled.
}
\label{fig:temp_color_comp_snpy}
\end{figure}

We also checked the broadband colors of the new template using the \texttt{SNooPy} model.
The broadband colors of the new spectral template and the light-curve model of \texttt{SNooPy} are in general agreement within 0.2~mag for all colors and $s_{BV}$ values (Figure~\ref{fig:temp_color_comp_snpy}).
The large discrepancies at the early phases may be due to the lack of observed data.
The general agreement gives assurance that the merging of the seven wavelength regions at the last step of the template construction preserves the broadband color information.
Note, however, that the spectral template should always be color-matched to the observed photometric colors of the SN~Ia in question during the light-curve fitting process.

Another test we did is to address the potential training bias toward SN~2012fr, the best-observed SN in the sample, with 25 spectra out of the total 331 sample spectra. 
Following the methodology of leave-one-out cross validation, we constructed another template without SN~2012fr following the same methodology to compare with the template built with the whole sample.
The two templates were color-corrected to have the matching broadband colors.
The maximum flux difference is around 6\% in the region near 0.85~$\mu$m, where SN~2012fr presents the detached high-velocity \ion{Ca}{2} feature at early times.
In general, the differences between the two templates are negligible, with a median flux difference less than 1\%.


\subsection{K-correction Uncertainties}
\label{sec:dkcorr}

We utilized the definitions of K-correction uncertainties from \citet{Hsiao2007} to evaluate the performance of the new template. 
K-corrections obtained using the template spectra ($K_{temp}$) were compared to those from observed spectra of our sample ($K_{obs}$) for different combinations of filters and redshifts.

The RMS and the mean of the K-correction differences were then taken as the statistical and systematic uncertainties, respectively.
The K-correction uncertainties were also compared to those computed using the Nugent \citep{Nugent2002} and Hsiao \citep{Hsiao2007} templates.
The computations were done using the \texttt{SNooPy} package, while the template spectra were color-corrected to match the broadband $YJH$ colors of the observed spectra before the computation.

Note that these uncertainty measurements were designed solely to evaluate how well the template can predict the spectral features of the observed spectra.
The broadband colors of the template were corrected to match those of the observed spectra, as it is common practice in light-curve fitting to correct the template to the observed colors.
These estimates do not include other effects, such as the uncertainty in the relative flux calibration discussed in Section~\ref{subsec:sample_distribution}.

In the most ideal case, one should assess the performance of the new template using a validation set of observed spectra that is completely independent of the template training set.
In the NIR, the sample size is severely restricted by the telluric region, as discussed in Section~\ref{sec:sample_preparation}.
There is currently no spectral sample with reliable telluric regions besides the FIRE sample presented here, and after inspecting the telluric corrections, our sample was reduced by roughly 1/3.
Given this difficulty, we chose to evaluate its performance using a cross-validation technique described below.

First, a subsample with reliable telluric regions was chosen to form a ``validation" pool. 
Next, we randomly selected 66 spectra from this pool to form the validation set, which constitutes 20\% of the entire sample.
The remaining 80\% of the entire sample was then used as the training set.
The training set was used to construct a template with the procedures outlined in Section~\ref{sec:methods}, while the validation set was used to estimate the K-correction uncertainties.
The process of splitting the sample, constructing the template, and estimating the K-correction uncertainties was repeated for 1{,}000 iterations.
The resulting uncertainty estimates converge rather quickly, generally after 50 iterations, while the 1{,}000 iterations yielded Gaussian-like distributions.

\begin{figure*}
\centering
\includegraphics[width=0.95\textwidth]{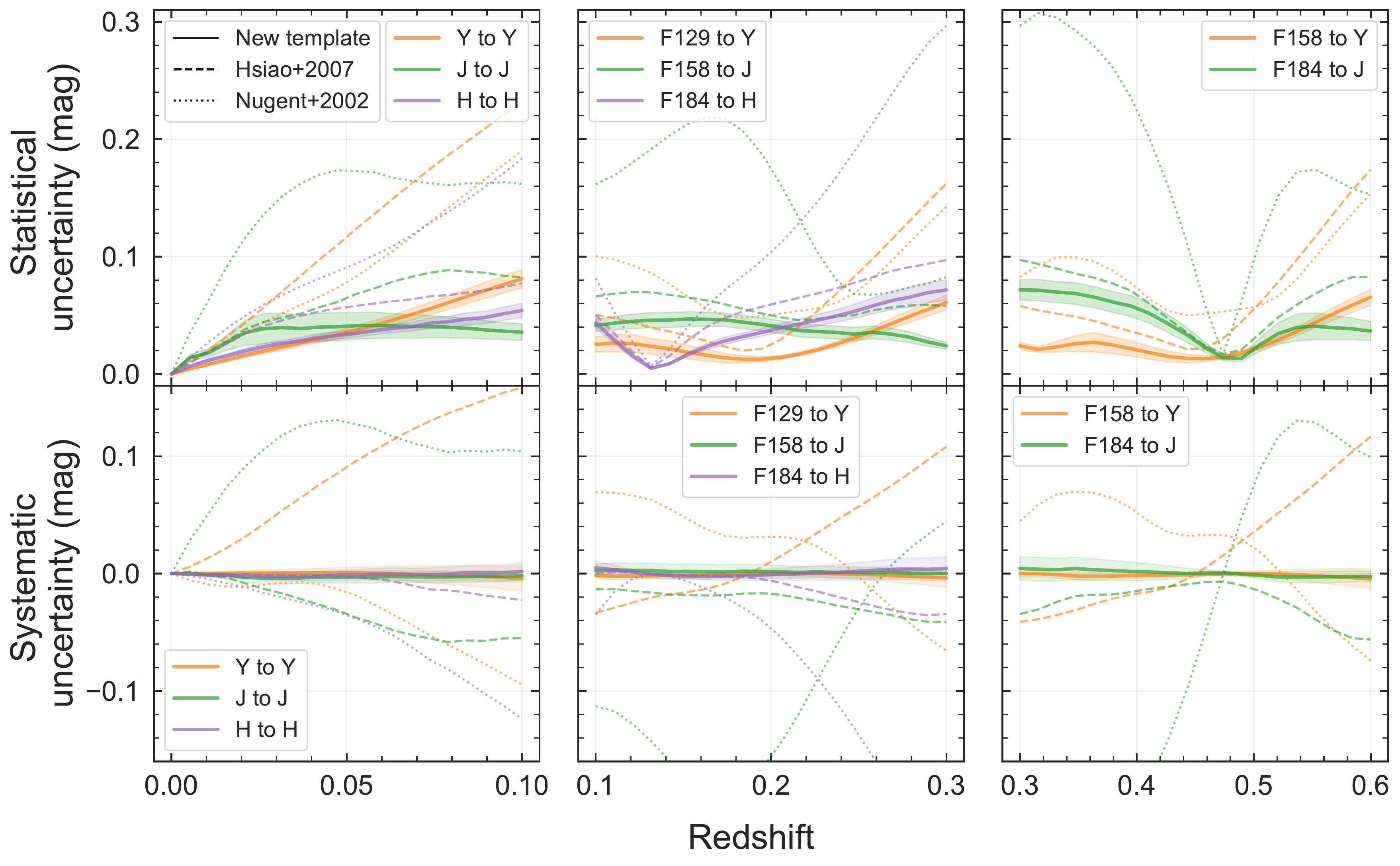}
\caption{
K-correction uncertainties of the new template compared to those of the Nugent \citep{Nugent2002} and Hsiao \citep{Hsiao2007} templates.
The RMS and the mean of the K-correction differences between the template and observed spectra are taken as the statistical (top panels) and the systematic (bottom panels) uncertainties.
The mean and standard deviation of the 1{,}000 iterations are shown as solid curves and shaded regions, respectively. (See text for details.)
The filters used are the $YJH$ scans of du Pont + RetroCam, as well as the proposed RST + WFI $F129$, $F158$, and $F184$ filters.
}
\label{fig:pca_kcorr_diff}
\end{figure*}

Figure~\ref{fig:pca_kcorr_diff} presents the K-correction uncertainties as a function of redshift for several cases of single-filter K-corrections at low redshifts and cross-filter K-corrections at high redshifts.
The mean and standard deviation of the 1{,}000 iterations are shown as solid curves and shaded regions, respectively.
The observer-frame filters for the high-redshift cases are the Wide Field Instrument (WFI) $F129$, $F159$, and $F184$ filters of the RST \citep{Hounsell2018}\footnote{\url{https://roman.gsfc.nasa.gov/science/WFI_technical.html}}.
The remaining filters are of RetroCam on the du Pont Telescope, measured by \citet{Rheault2014}.

As shown in the upper panels of Figure~\ref{fig:pca_kcorr_diff}, the statistical K-correction uncertainties of the new template are significantly reduced compared to the previous templates. 
For example, at $z=0.05$, improvements of 0.09, 0.03, 0.02~mag in $Y$, $J$, and $H$ band, respectively, can be gained by switching from the Hsiao template to the new one.
Note that the result of \citet{Boldt2014} that the smallest uncertainties can be found in the $Y$ band is not reproduced here.
From a larger sample, we found that all $YJH$ bands yield similar uncertainties.
For cross-filter K-corrections, a minimum would be present at a redshift where the observed and rest-frame filters are aligned \citep{Kim1996}.
On average, there is a 0.02~mag improvement in statistical uncertainty compared to the Hsiao template in cross-filter K-corrections.

The lower panels of Figure~\ref{fig:pca_kcorr_diff} demonstrate that the systematic K-correction uncertainties are essentially eliminated by adopting the new template. 
This is true for all the redshifts and filter combinations considered here, and represents as 0.1~mag improvements in some cases.
The improvements in the $Y$-band are the most evident.
Since the broad-band colors of the template spectra were matched to the observed spectra before the comparison, these uncertainty estimates indicate the ability of the template to predict the spectral feature profile shapes.

\begin{figure*}
\centering
\includegraphics[width=0.95\textwidth]{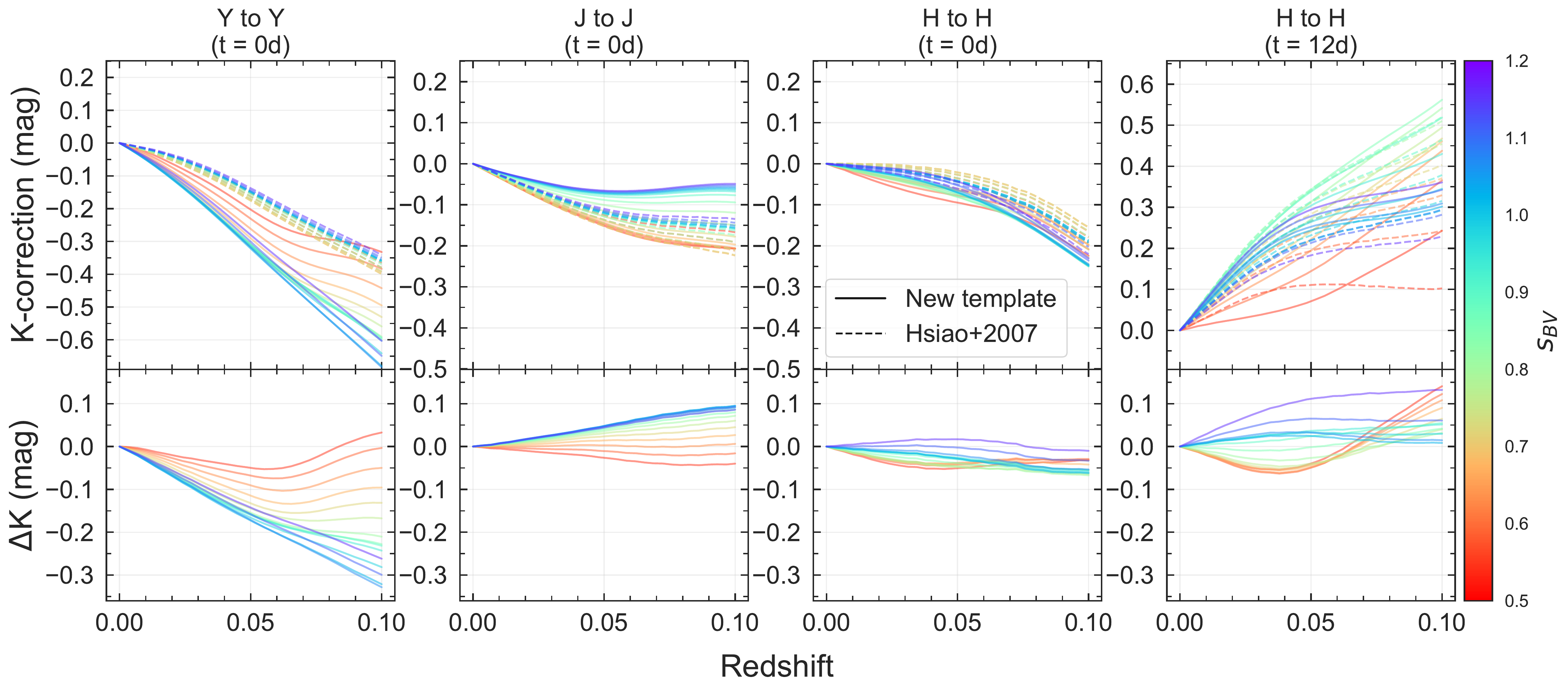}
\caption{The K-corrections of our new template compared to the previous Hsiao template \citep{Hsiao2007,Hsiao2009} at maximum light $T_{\text{max}}^{B}$.
Both templates are color corrected to match the \texttt{SNooPy} light-curve templates, which is a function of $s_{BV}$.
The bottom panels plot the K-correction differences between the two templates: $\Delta$K = K(New template) $-$ K(Hsiao).
Note that the variation in K-corrections of the new template is caused by both the diversity of the temporal spectral features and the broadband colors, while the variation of the Hsiao template is only reflecting the color variation of the \texttt{SNooPy} light-curve templates in $s_{BV}$.
The K-correction differences can be substantial, such as for slow decliners at redshift $\sim$0.1, the $\Delta$K in Y band is as large as 0.3~mag.
}
\label{fig:kcorr_diff_at_Bmax}
\end{figure*}

\begin{figure}[tbh!]
\centering
\includegraphics[width=0.98\columnwidth]{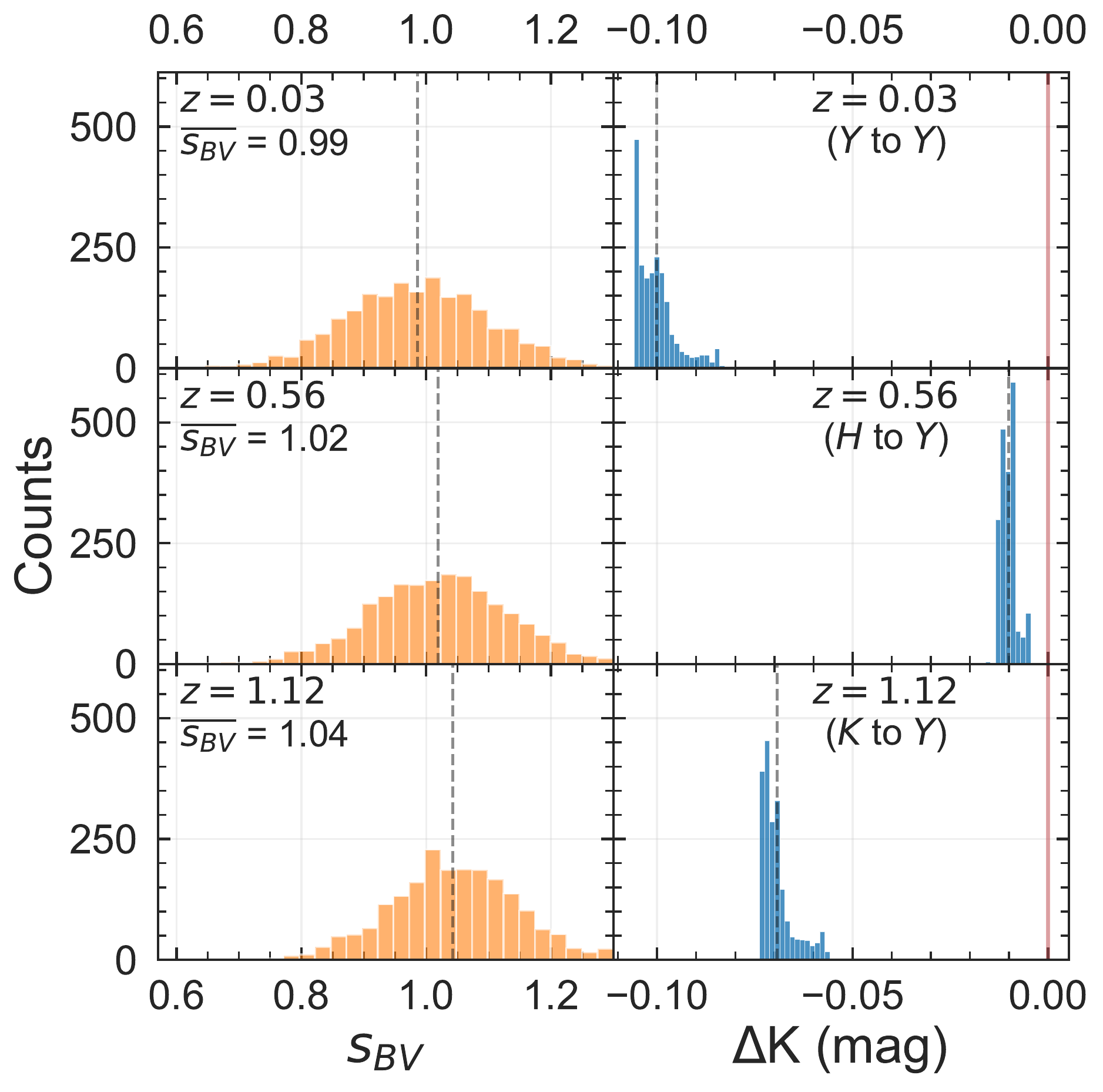}
\caption{
Simulating the effects of an evolving $s_{BV}$ distribution as a function of redshift.
\textit{Left panels:} Three redshift bins are considered: one local and two at high redshifts.
The $s_{BV}$ distributions are assumed to be Gaussian with the mean converted from the stretch values taken from \citet{Howell2007}.
The vertical dashed lines mark the mean of the distribution in each bin.
\textit{Right panels:}
The distributions of the resulting peak-magnitude K-correction differences are shown, assuming the distribution of $s_{BV}$ in the corresponding left panel.
The definition of $\Delta$K is the same as in Figure~\ref{fig:kcorr_diff_at_Bmax}.
The vertical dashed lines mark the mean of the distribution in each panel.
The red solid vertical line marks the position of $\Delta$K = 0.
}
\label{fig:kcorr_diff_at_3z}
\end{figure}

\subsection{Impact on Cosmology}
\label{sec:cosmology}

Here, we explore the impact of adopting the new SN~Ia NIR spectral template on cosmological analyses, both at low (e.g., Hubble constant measurements) and high (e.g., dark energy measurements) redshifts.

We first examine the impact on the NIR peak magnitude of adopting the new template on the low-redshift end in single-filter K-corrections.
The results are presented in comparison to the Hsiao template in  Figure~\ref{fig:kcorr_diff_at_Bmax}.
Note that we chose to make the calculations at $B$-band maximum for convenience, and the analysis yielded a rough estimate of potential magnitude change around peak magnitude.
Both templates are scaled and color-corrected to match the $YJH$ magnitudes of the \texttt{SNooPy} light-curve template, which is a function of $s_{BV}$.
Recall that the Hsiao template at peak is a single spectrum and does not account for the variation of spectral features.
The spread of roughly 0.1~mag in K-corrections at $z=0.1$ results entirely from color corrections.

On the other hand, the new template at peak is a series of spectra that are a function of $s_{BV}$.
Figure~\ref{fig:kcorr_diff_at_Bmax} shows the difference between the two templates in terms of the K-correction as a function of redshift.
It essentially reflects the fact that the diversity is not only in the broad-band colors but also in the spectral features.
Not accounting for the variation of spectral features as a function of $s_{BV}$ can cause errors in K-corrections as large as 0.3~mag in $Y$ and 0.1~mag in $J$ at $z=0.1$.
Also note that the two templates show the largest difference in slow decliners (high $s_{BV}$ values).
The prominent $H$-band feature is at maximum strength around 12~days after the maximum.
Both slow and fast decliners show K-correction differences as large as 0.1~mag at $z = 0.1$.

For cosmological analyses using distant SNe~Ia, the redshift evolution of the SN~Ia population is an important factor to consider. 
Based on previous high-redshift surveys, the SN~Ia population at high redshifts tend to be slower decliners and have higher optical light-curve stretch values on average \citep[e.g.,][]{Howell2007,Nicolas2021}.
\textit{This would introduce an additional systematic shift in the K-corrections if the template SEDs do not reflect the spectral feature variations with light-curve shape.}

We simulate such a scenario, considering a shifting $s_{BV}$ distribution at three different redshifts.
In the left panels of Figure~\ref{fig:kcorr_diff_at_3z}, three populations of SNe~Ia are presented at three different redshifts, simulating the distribution of SN~Ia light-curve shapes observed locally and at high redshifts \citep{Howell2007, Riess2007}.
The mean of the $s_{BV}$ distributions shifts by as much as 5\% from the local sample to $z=1.12$.
We then calculated K-corrections to the rest-frame $Y$ band and examined the difference between the new and Hsiao templates at $B$-band maximum.

The resulting K-correction differences from the corresponding $s_{BV}$ distribution are presented in the right panels of Figure~\ref{fig:kcorr_diff_at_3z}.
The observer-frame filters for the two high-redshift bins, $H$ and $K$, align well with the rest-frame $Y$ band at each redshift.
The effective wavelength of the de-redshifted $H$ ($z=$0.56) and $K$ ($z$=1.12) bands match that of the rest-frame $Y$ band within 0.02~$\mu$m.
Even so, the K-correction differences are systematic and significant, with the mean of 0.10, 0.01, 0.07~mag at $z=0.03, 0.56, 1.12$, respectively.
The results again highlight the importance of accounting for the spectral feature variations as a function of light-curve shape \citep[e.g.,][]{Jones2022}.


\section{Conclusion} \label{sec:conclusion}

We present 339 NIR spectra of 98 SNe~Ia observed with Baade + FIRE by the CSP-II \citep{Phillips2019,Hsiao2019}, the largest and the most homogeneous NIR spectral sample of SNe~Ia to date.
Among those, 331 spectra of 94 SNe~Ia were used to construct the most accurate NIR spectral template as a function of a light-curve-shape parameter.
The spectra maintain a spectrophotometric accuracy on the level of $10-20$\%, allowing for examination of the broad-band colors along with the spectral features.
The telluric regions are crucial when we study SNe~Ia at a range of redshifts.
Thankfully, the high-throughput nature of the FIRE NIR spectrograph enables consistent and reliable telluric corrections for $\sim$70\% of the sample.

The aim is to obtain an accurate description of the SN~Ia SED in the NIR as a function of phase and $s_{BV}$.
To achieve this, we first utilized PCA to reduce the dimensionality of the data set, then used GPR to model the hypersurface in phase and $s_{BV}$ space.
The wavelength grid was divided into seven regions in order to fully utilize the sample and to avoid the PCA results being dominated by spurious pixels or intrinsically high flux.
The GPR $R^2$ score has proven to be an efficient metric for selecting which PCs to include for the reconstruction of the SED, capturing the most data variation while using the least number of PCs.
The GPR approach also allowed the estimation of the flux uncertainty for each template SED.

The new template successfully captures the diversity of the broad-band colors and spectral features as a function of phase and $s_{BV}$.
The hallmark SN~Ia NIR features, such as \ion{Mg}{2} at early times and the $H$-band break past maximum, are recreated in the template SED.
In NIR, SNe~Ia have more spectral variation than in the optical that cannot be simply described by the varying speed of the evolution for SNe~Ia with a range of decline rates.
The template is also able to predict the SED shapes of spectra that are not in the training sample and produce photometric colors consistent with the \texttt{SNooPy} light-curve model. 
Using the cross-validation method, randomly splitting the sample into training and testing sets, a drastic decrease in the K-correction uncertainty is shown compared to previous templates.
The new template essentially eliminates any systematic K-correction uncertainties with a 90\% improvement compared to the Hsiao template \citep{Hsiao2009}.
Simulations of low- and high-redshift cosmological analyses illustrate problems with assuming a fixed template that does not vary with $s_{BV}$.

The new spectral template will be implemented in \texttt{SNooPy}\footnote{\url{https://github.com/obscode/snpy}} \citep{Burns2011} and is available on GitHub\footnote{\url{https://github.com/DeerWhale/BYOST}} along with the source code.
It can also be adopted as the baseline SED for other light-curve fitters, such as \texttt{BayeSN} \citep{Mandel2009,Mandel2011,Mandel2022}.
As the template fully captures the spectral behaviors of normal SNe~Ia, the PCs can be used to identify peculiar features, interesting physics, and possible secondary parameters that could improve their cosmological utility.
It will, for example, be revealing to cross-compare these NIR-based PCs with the 3-parameter nonlinear parameterization of the optical spectral behaviors of SNe~Ia developed by \citet{Boone2021b} from a similarly untargeted sample of 173 SNe, which leads to a much improved (sigma $\sim$0.08~mag) dispersion in cosmological distance measurements \citep{Boone2021c}.
The machine-learning techniques developed here will also have a wide range of applications for detecting and modeling the variation in a large and multidimensional data set.
Finally, understanding the NIR spectral diversity represents a crucial step forward for future SN~Ia cosmological experiments, such as that to be conducted on the Nancy Grace Roman Space Telescope.

\section*{Acknowledgments}
The authors would like to thank the referee for providing helpful comments.
We thank the technical, scientific staff, and support astronomers of the Las Campanas Observatory for all the support over the years.
The CSP-II has been supported by NSF grants AST-1008343, AST-1613426, AST-1613455, AST-1613472, and the Danish Agency for Science and Technology and Innovation through a Sapere Aude Level 2 grant (PI: M.S.).
We thank Prof.~Sean Dobbs and Dr.~Daniel Lersch for the useful discussion. 
The Aarhus supernova group is funded by an Experiment grant (\# 28021) from the Villum FONDEN, and by a project 1 grant (\#8021-00170B) from the Independent Research Fund Denmark (IRFD).
C.A. acknowledges STSci grant JWST-GO-02114.032-A.
C.G. is supported by a research grant (25501) by the VILLUM FONDEN.
E.B. acknowledges NASA Grant 80NSSC20K0538 and STSci grant JWST-GO-02114.032-A.
L.G. acknowledges financial support from the Spanish Ministerio de Ciencia e Innovaci\'on (MCIN), the Agencia Estatal de Investigaci\'on (AEI) 10.13039/501100011033, and the European Social Fund (ESF) "Investing in your future" under the 2019 Ram\'on y Cajal program RYC2019-027683-I and the PID2020-115253GA-I00 HOSTFLOWS project, from Centro Superior de Investigaciones Cient\'ificas (CSIC) under the PIE project 20215AT016, and the program Unidad de Excelencia Mar\'ia de Maeztu CEX2020-001058-M.
S.G.G. acknowledges support by FCT under Project CRISP PTDC/FIS-AST-31546/2017 and Project~No.~UIDB/00099/2020.

\facilities{Magellan Baade Telescope (FIRE), Swope Telescope}

\software{Astropy \citep{astropy:2013,astropy:2018}, Pandas \citep{pandas_paper,pandas_software}, PyTorch \citep{PyTorch}, scikit-learn \citep{scikit-learn}, SNooPy \citep{Burns2011}.}

\bibliographystyle{aasjournal}
{\footnotesize
\bibliography{ref}}


\appendix
\restartappendixnumbering 
\renewcommand{\thefigure}{A\arabic{figure}}
\renewcommand{\theHfigure}{A\arabic{figure}}

\section{NIR spectra sample tables} \label{appendix:NIR spec table}
In this work, we publish 339 NIR spectra of 98 SNe~Ia obtained by the CSP-II that passed the first two selection criteria outlined in Section~\ref{subsec: sample selection}.
Table~\ref{tab:SNe_and_spec} presents the information of the SNe~Ia and spectra.
Note that the $T_{\text{max}}^{B}$ and the $s_{BV}$ of the sample SNe were obtained by fitting multiband CSP light curves with \texttt{SNooPy}.
The phases are relative to $T_{\text{max}}^{B}$.
The spectral data and the information in this table are available in the CSP website (\url{https://csp.obs.carnegiescience.edu/data}).

\startlongtable


\restartappendixnumbering 
\renewcommand{\thefigure}{B\arabic{figure}}
\renewcommand{\theHfigure}{B\arabic{figure}}

\section{Alternative Method: Conditional Variational Autoencoder} \label{sec:cVAE}

\begin{figure*}
\centering
\includegraphics[width=0.98\textwidth]{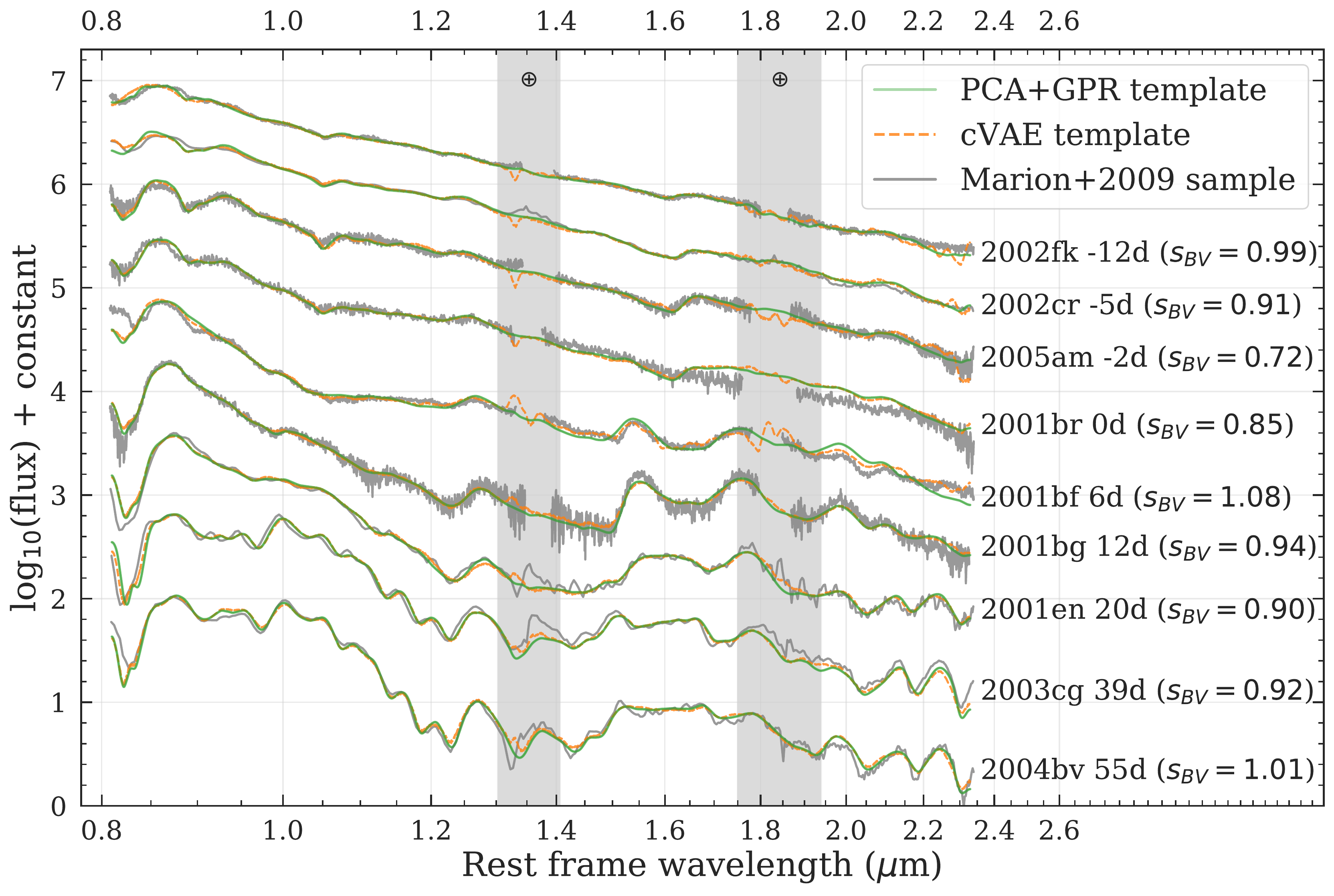}
\caption{
Comparisons between templates constructed using the current PCA+GPR method (green solid lines) and the cVAE method (orange dotted lines).
Selected observed spectra from \citet{Marion2009}, which are not included in the training process of either template, are also presented in the background for comparison (gray solid lines).
All comparison spectra are color-corrected to match the broadband colors.
Regions of strongest telluric absorptions are marked with vertical gray bands.}
\label{fig:cVAE_temp_comp}
\end{figure*}

As we mentioned in Section~\ref{sec:methods}, the current PCA+GPR method has some limitations.
Here we propose an alternative machine-learning method, using extensions of the autoencoder network, that could circumvent the drawbacks in the current approach.

Autoencoders have been employed in astronomy with rising popularity in recent years \citep[e.g.][]{Villar2020,Boone2021_ParSNIP,Gabbard2022}. 
An autoencoder is a type of neural network architecture that couples an encoder neural network and a decoder neural network, with some reduction in dimensionality at their connection point. 
This reduced space is known as the latent space of the network, and can be seen as a nonlinear subspace, in comparison with the linearly transformed subspace that is formed from PCA. 
The dimensionality of the latent space is akin to the choice of the number of PC dimensions used for template generation. 
The network is forced to reconstruct the inputs utilizing the fewest dimensions while incurring some reconstruction loss that can be minimized.

A variational autoencoder \citep[VAE;][]{vae} extends its predecessor by encoding for random variables. 
Typically, this is chosen to be a vector of means and variances for a set of Gaussian random variables. Decoding random samples from the latent space gives access to the probability density function of generated templates. 
An additional regularization (Kullback-Leibler divergence) term is added to the loss function of the standard VAE architecture that attempts to maintain some level of continuity in the latent space by guiding the embeddings to be more Gaussian distributed. 
This makes template generation smoother. Furthermore, an extension to the VAE architecture can be made by conditioning the encoder and decoder on external variables, e.g., phase and $s_{BV}$. 
This addition is known as a conditional VAE \citep[cVAE;][]{cvae} and allows us to output template spectra for given phase and $s_{BV}$ inputs continuously, without requiring an additional GPR step.

Here, a cVAE for template generation is implemented in the \texttt{pyTorch} \citep{PyTorch} framework, and its performance will be analyzed.
The encoder is constructed as a feed-forward neural network consisting of four layers containing 3673, 128, 32, and 5 nodes, respectively. 
Rectified linear units are used as activations. 
The decoder neural network, in this case, mirrors the encoder with the exception that the first node contains 5+2 nodes (the +2 is reserved for conditioning the decoder on the template parameters). 
The cVAE is optimized using the Adam optimizer \citep{Adam_optimizer} with a learning rate of $10^{-3}$ and a batch size of 16.
The model is trained for 500 iterations.
Under the current settings, the computation time of cVAE is 4 times longer than the PCA+GPR method, at around 40~minutes and 13~minutes, respectively.
However, the time would depend on the hardware and the choice of iterations.

Figure~\ref{fig:cVAE_temp_comp} shows a comparison of the templates generated by PCA+GPR and the proposed cVAE method.
The current PCA+GPR method still provides the best results, with a median $\chi^2/dof$ of 1.43 while compared to observed spectra (see Section~\ref{sec: PC selection}). 
The templates constructed with cVAE method yield a median $\chi^2/dof$ of 1.76. 
The current data set might not be large enough to train the cVAE to adequately separate the signal from the noise, especially in the telluric regions and the noisy edge of the $K$ band. 
Further exploration, such as refining selection criteria and training for more iterations, is required to further investigate the causes of deficiencies in these regions.

However, there are caveats in this comparison of the resulting templates. 
The input data of cVAE and the PCA+GPR method do not have the same structure: the input data of cVAE is \textit{not} split into wavelength regions and the input data is normalized in both mean and standard deviation, whereas the input data for PCA is only normalized by mean in each column.
Additionally, only $\sim$70\% of the sample is included for training in the cVAE approach due to the limitation of the reliable telluric regions, which essentially reduces the training size by 30\% in the $zYJH$ and 20\% in the $K$ band compared to the PCA+GPR method.

The proposed cVAE architecture requires fixed wavelength grid as an input, similar to PCA. A useful extension to this would be to support sequence inputs of varying length, thereby skipping any implicit biases that can be introduced by the interpolation preprocessing step. One clear and natural pathway forward would be to incorporate long short-term memory (LSTM) or recurrent neural network (RNN) nodes. Here, a single node takes in an entire sequence, element by element, and contains an internal state that allows information from previous elements of a sequence to affect the output for the next element. Using a sequence approach is also beneficial in that different regions can be ignored, e.g., wavelength regions that have poor telluric corrections.

In summary, the cVAE approach has the following potential advantages over PCA+GPR: it can naturally deal with irregular/missing data and unifies the two-step PCA+GPR procedure allowing the template generation and dimensionality reduction steps to communicate. 
However, it appears that a larger data set is needed to better harvest the potential of this method.

\restartappendixnumbering 
\renewcommand{\thefigure}{C\arabic{figure}}
\renewcommand{\theHfigure}{C\arabic{figure}}

\section{Gaussian Process Regression Kernel parameters}\label{appendix:GP kernal parameters}

In this section, we present the details of the GPR kernel set up and the optimal hyperparameters, as well as a demonstration of how the length scales of the RBF kernel affect the template spectra.

In section~\ref{sec:GPR}, GPR was used to establish the average dependence of the PC projection values 
on phase and decline-rate, $p_i\left(phase,s_{BV}\right)$. With $p_i\left(phase,s_{BV}\right)$, one 
can ``look up" the PC projections and construct the average SED of a SN~Ia for any particular value of $s_{BV}$ and phase. 
GPR has the advantage of being non-parametric, in the sense that no particular functional form is chosen
to represent $p_i\left(phase,s_{BV}\right)$. 
Rather, GPR determines a mean function which is constrained by the data and a covariance function, called the kernel, that controls the amplitude and scale over which this function can vary.

For the kernel, we choose the sum of a RBF covariance function and a white noise term, which has 4 
free hyperparameters and can be represented as the following: 
\begin{equation} \label{Eq:GPR_kernel}
kernel = C \times \rm{RBF}(\ell_{phase},\ell_{s_{BV}}) + \rm{W}(\sigma).
\end{equation}
C is a constant, representing the amplitude of the RBF kernel, $\ell_{phase}$ and $\ell_{s_{BV}}$ are the length scales of the RBF kernel for phase and $s_{BV}$, and $\sigma$ is the noise level of the white noise kernel, $W$.
The values of these hyperparameters are determined by {\tt scikit-learn} by maximizing the log-marginal likelihood.
All 4 parameters are given uniform priors.

The shape of the probability hypersurface is a complicated one, with obvious degeneracies. The most
severe occurs when $\sigma$ becomes larger than the variations due to $s_{BV}$ and/or phase.
In this case, the RBF scales ($\ell_{phase}$ and/or $\ell_{s_{BV}}$) no longer matter and therefore
do not converge to fixed values. 
In these cases, the resulting $p_i\left(phase,s_{BV}\right)$ are
simply constant with respect to phase and/or $s_{BV}$.

\begin{figure*}
\centering
\includegraphics[width=0.98\textwidth]{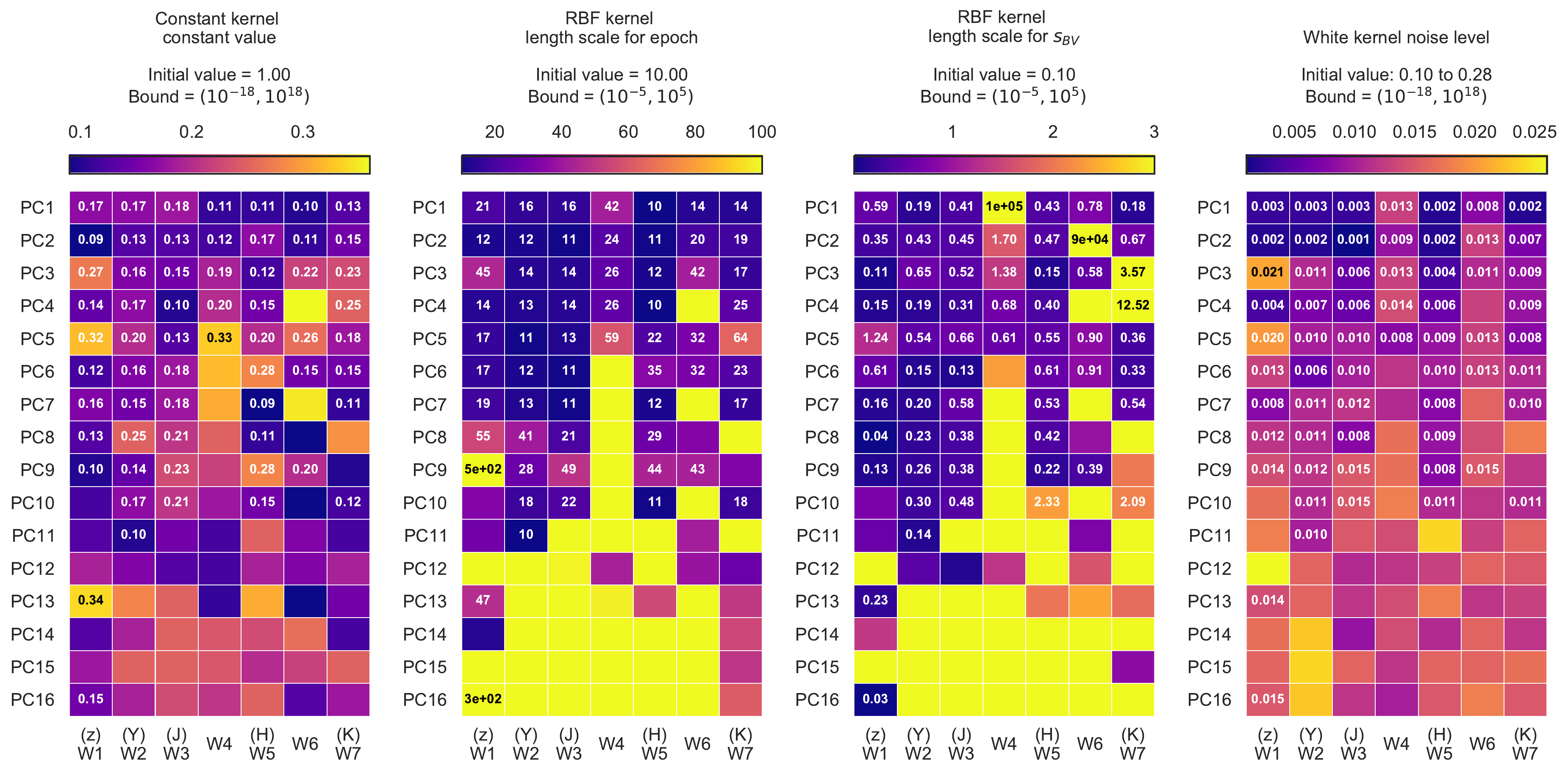}
\caption{
The heatmaps of the optimal hyperparameters of GPRs.
From left to right are the constant (C), RBF kernel length scale for phase ($\ell_{phase}$), RBF kernel length scale for $s_{BV}$ ($\ell_{s_{BV}}$), and White kernel noise level ($\sigma$), respectively.
The selected PCs for the template constructions with GPR R$^2 \ge 0.2$ (see Section~\ref{sec: PC selection}) are indicated with labels.
}
\label{fig:GPR_optimized_hyperparameters}
\end{figure*}

The optimal hyperparameters of all GPRs are presented in Figure~\ref{fig:GPR_optimized_hyperparameters}.
Around half of the GPRs, especially those of the lower rank PCs, do not have converged length scales, in
which case we set the scale equal to 10$^5$, effectively making the GPR independent of that parameter.
However, our PC selection strategy (GPR R$^2 \ge 0.2$; see Section~\ref{sec: PC selection}) was able to screen out the vast majority of GPR fits that have these convergence issues.  
The only selected PC that has not fully converged is PC1 of W4 (telluric) region, which has a $\ell_{s_{BV}} = 10^5$.
In general, for the templates presented in this work, the median optimal length scales of the selected PCs are 18 days and 0.43 for phase and $s_{BV}$, respectively. 
The optimal length scales for phase are similar to the best-fit length scale for the 
\texttt{SNooPy} light-curve template ($\sim$30~days), which were also constructed with GPR.
PCs that have relatively short optimal length scales may indicate spectral features that evolve faster than the photometry, such as those due to ionization changes.

While our analysis provides optimal values for the four hyperparameters, there are associated uncertainties in their values. 
It is therefore worth considering the effects of varying each hyperparameter with respect to the constructed SN~Ia SED. 
In other words, how do uncertainties in the hyperparameters propagate to uncertainties in the final SN~Ia SED?
Since the phase and $s_{BV}$ are the two input parameters for the template, the temporal spectra are likely to vary with their RBF length scales the most. 
To illustrate how each of the length scales affect the final template, we keep one length scale at 
the optimal value, while varying the other by 1$\sigma$ Median Absolute Deviation (MAD).
The MAD of $\ell_{phase}$ and $\ell_{s_{BV}}$ among selected PCs are 6.6 and 0.22, respectively.

\begin{figure*}
\centering
\subfigure{\includegraphics[width=0.4\textwidth]{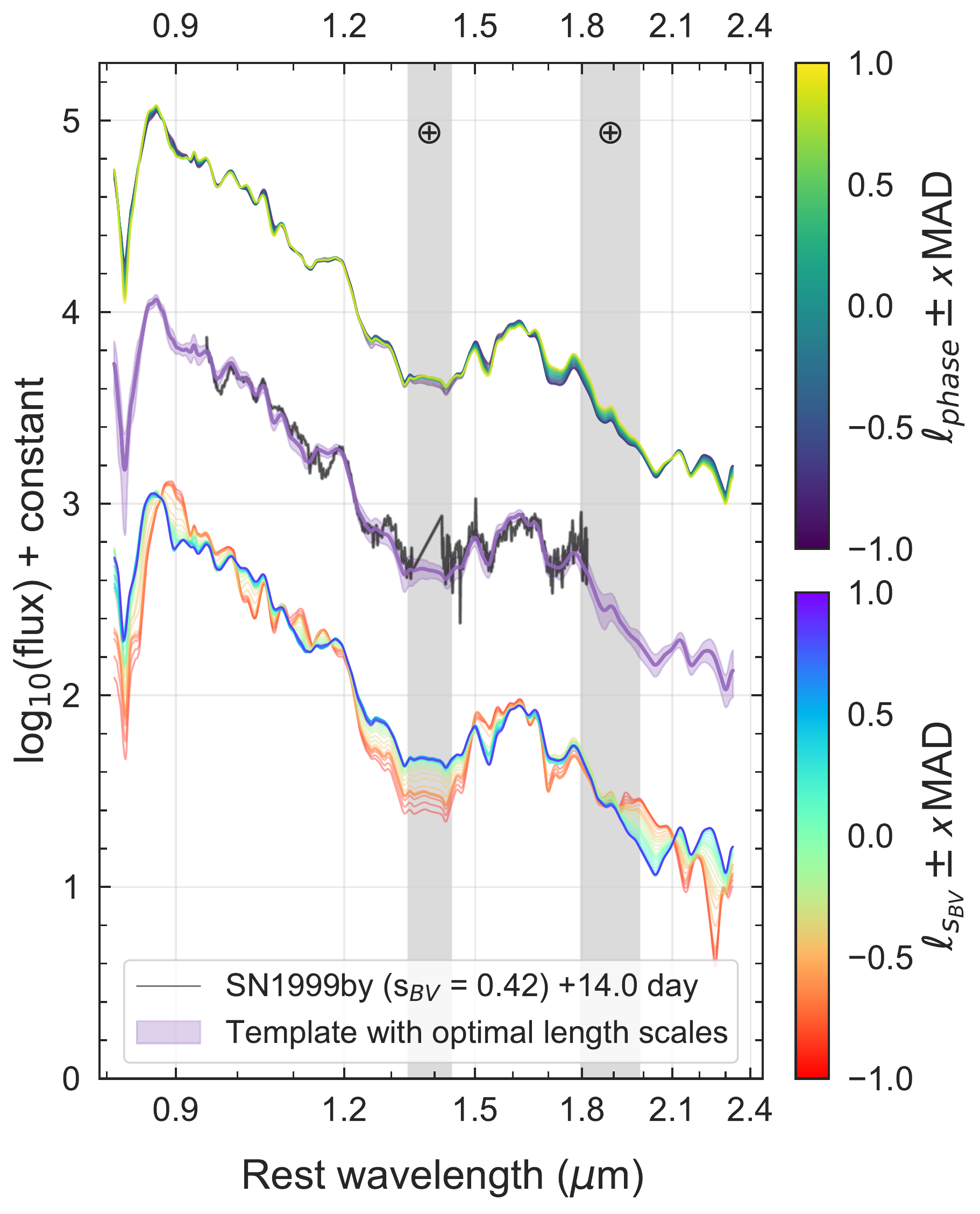}}
\subfigure{\includegraphics[width=0.4\textwidth]{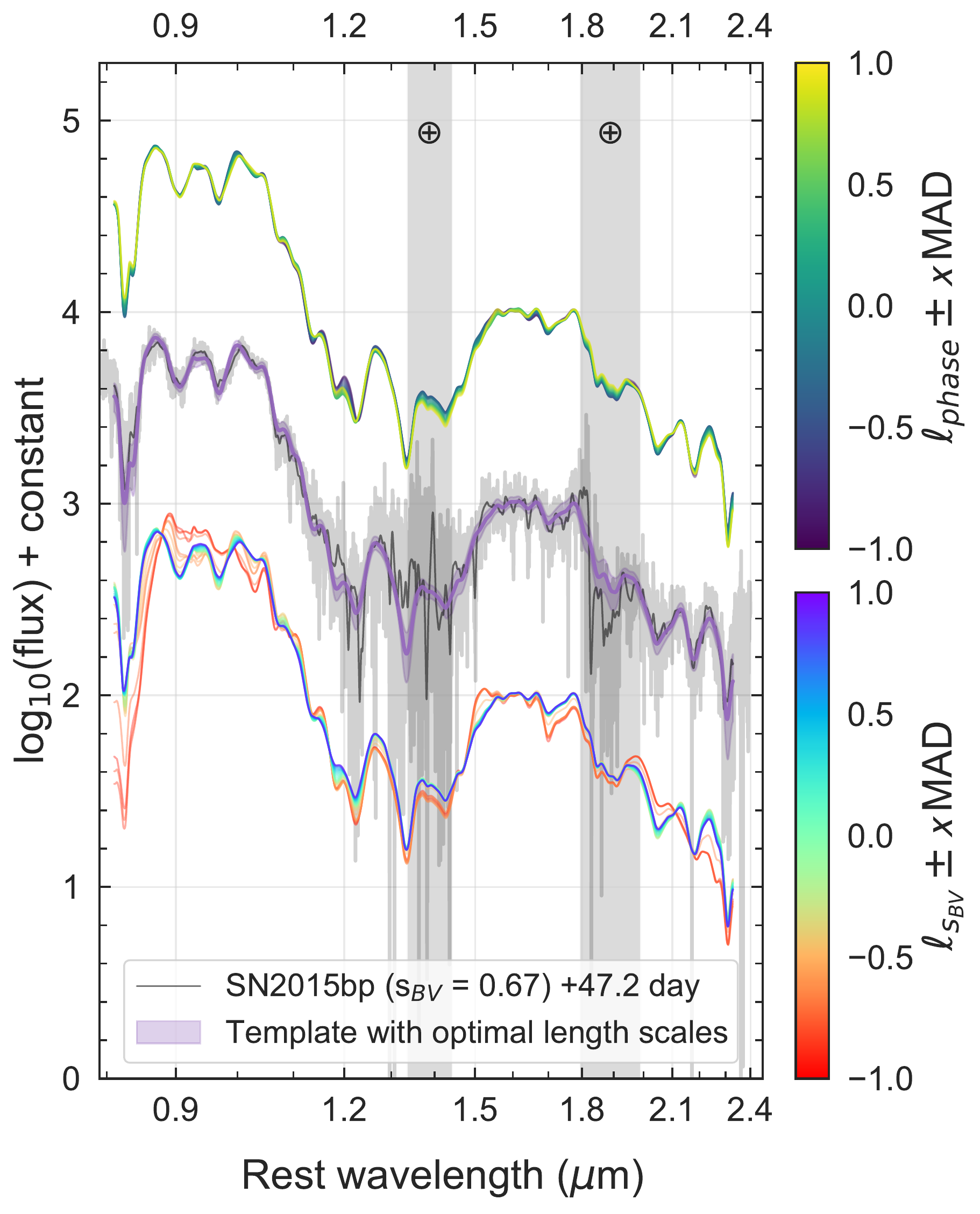}}
\caption{
Comparison between the test templates with varying length scales, the observed spectra, and the optimal template.
The phase and $s_{BV}$ length scales are varied by 1$\sigma$ median absolute deviation and are plotted on the top and bottom of the observed spectrum, respectively.
The template is less robust against varying length scales in regions of the parameter space that are not well-modeled, such as $s_{BV} = 0.42$ (left panel).
The right panel shows the same comparison but at $s_{BV} = 0.67$ and phase $= +47.2$~days, which is on the border of extrapolation. 
Note that the spectra are color-matched in this comparison. 
}
\label{fig:temp_vary_lengthscales}
\end{figure*}

The resulting test templates are shown in Figure 
\ref{fig:temp_vary_lengthscales}, where variations in $\ell_{phase}$ are plotted above the data as a sequence from blue to yellow lines and variations in $\ell_{s_{BV}}$ are plotted below as a sequence from red to blue lines.
On the left, we plot SN~1999by at $+14.0$ days, which has a decline rate $s_{BV}=0.42$ and on the right, we plot SN~2015 at $+47.2$ days, which has a decline rate $s_{BV}=0.67$. 
These were chosen to represent points in the phase-$s_{BV}$ space that are sparsely sampled by the training data (see
Figure~\ref{fig:counts_phase_sBV_grid}). 
SN~1999by, being on the low edge of the $s_{BV}$ distribution, shows a larger variation when varying $\ell_{s_{BV}}$ than in the case of SN2015bp, which resides in a better-sampled part of the $s_{BV}$ distribution.

It should be emphasized that the ``errors" seen in \ref{fig:temp_vary_lengthscales} are exaggerated since the real variations in $\ell_{phase}$ and $\ell_{s_{BV}}$ due to uncertainties are lower (on the order of 0.2~days and 0.1, respectively). Nevertheless, it illustrates that improvements can be made by adding more training data in those areas that are sparsely sampled in $s_{BV}$ and phase, particularly fast-declining SNe at late phases.

\end{document}